# Divergent Solid-state Conversion Pathways in Evaporated All-perovskite Tandem Solar Cells

Huagui Lai[1,2], Amber Wright[1], Nick Huber[1], Niels Uythoven[1], Federico De Giorgi[1], Jincheng Luo[1], Tristan Sachsenweger[2], Sunil B. Shivarudraiah[3], Chih-Jen Shih[3], Wolfgang Tress[2,*], Fan Fu[1,*]

[1]Laboratory for Thin Films and Photovoltaics, Empa – Swiss Federal Laboratories for Materials Science and Technology, Dübendorf 8600, Switzerland

[2]Institute of Computational Physics, Zurich University of Applied Sciences, Winterthur 8400, Switzerland

[3]Institute for Chemical and Bioengineering, ETH Zürich, Zürich 8093, Switzerland

Correspondence and requests for materials should be addressed to F.F. (fan.fu@empa.ch) or W.T. (trew@zhaw.ch).

## Abstract

Sequential thermal evaporation (sTE) is emerging as a solvent-free route to high-quality mid-bandgap perovskites, but its extension to mixed-halide wide-bandgap (WBG) and Sn–Pb narrow-bandgap (NBG) absorbers for all-perovskite tandem solar cells (TSCs) remains limited by an incomplete understanding of solid-state conversion. Here, using time-sliced ex situ analysis, we reveal divergent solid-state conversion mechanisms in sequentially evaporated WBG and NBG precursor stacks. In WBG stacks, formamidinium (FA)-containing species penetrate the inorganic template and Br/I redistribution precedes substantial three-dimensional perovskite formation. The photoactive phase then crystallizes from a chemically mixed reservoir, and absolute $PbBr_2$ thickness, rather than nominal $PbBr_2/PbI_2$ ratio, determines the final bandgap. In NBG stacks, by contrast, an early Pb-rich perovskite phase forms upon formamidinium iodide deposition, restricting further FA penetration into the buried $SnI_2$ precursor. Subsequent annealing promotes rapid lattice reorganization faster than Sn/Pb interdiffusion, leaving vertical compositional gradients. Guided by these insights, we develop sTE absorbers with bandgaps spanning 1.26–1.96 eV and demonstrate the first evaporated all-perovskite TSC, reaching a power conversion efficiency of 19.2%. Encapsulated tandems retain on average 80% of their initial efficiency after 1,200 h at 65 °C (ISOS-D-2). These results establish bandgap-specific control of solid-state conversion as a design principle for sequentially evaporated perovskite tandem photovoltaics.

## Main

All-perovskite tandem solar cells (TSCs) offer a route beyond the single-junction efficiency limit by pairing a mixed-halide wide-bandgap (WBG, ~1.8 eV) absorber with a Sn–Pb narrow-bandgap (NBG, ~1.2 eV) absorber. State-of-the-art all-perovskite tandems now exceed 30% efficiency, but rely predominantly on solution processing.[1,2] Such fabrication typically involves toxic solvents and antisolvent-quenched crystallization, and each wet-processing step can perturb the underlying subcell.[3] Moreover, neither spin coating nor antisolvent quenching translates readily to large-area, flexible or textured substrates.[4,5] Vacuum deposition offers a solvent-free alternative with precise thickness control, conformal coating and compatibility with industrial thin-film manufacturing.[6-8] Among vacuum routes, sequential thermal evaporation (sTE) is particularly attractive because inorganic and organic salts are deposited separately, allowing precursor amounts to be defined by layer thickness while avoiding the flux-matching and cross-contamination constraints of multi-source co-evaporation.[9]

This processing simplicity shifts the central challenge from stoichiometry control during deposition to absorber formation during post-deposition solid-state conversion. In sTE, the absorber is not deposited directly as a perovskite; instead, its formation begins with a chemically layered precursor stack and proceeds through thermally driven solid-state mass transport, reaction and crystallization. Previous sTE studies have shown that precursor ordering and annealing conditions strongly affect mid-bandgap (MBG) Pb-only absorbers,[9-11] and that precursor components and thermal treatment influence phase evolution and film crystallization in mixed-halide and all-inorganic WBG stacks.[12-14] For NBG Sn–Pb absorbers, existing studies on sTE have largely relied on pre-alloyed Sn–Pb iodide sources,[15,16] leaving the conversion of separately stacked $SnI_2/PbI_2$ precursors largely unresolved. These previous studies establish the importance of precursor architecture and annealing conditions but do not reveal how an initially layered stack evolves into a photoactive absorber. Specifically, it remains unclear how formamidinium (FA)-containing species penetrate different inorganic layers, how Br/I and Sn/Pb redistribution processes couple to intermediate-phase evolution and crystallization, and when the final absorber composition and bandgap are established.[17-19] This knowledge gap limits the rational design of sequentially evaporated WBG and NBG absorbers for efficient all-perovskite tandems.

Here we resolve how sequentially evaporated WBG and NBG precursor stacks convert into photoactive absorbers. Using a time-sliced ex situ strategy to capture essential characteristic stages of solid-state conversion, we trace the temporal coupling between precursor redistribution, intermediate-phase evolution and perovskite phase formation. We find that the two stacks follow divergent solid-state conversion pathways, each governed by distinct kinetic bottlenecks. In WBG stacks, halide redistribution outpaces crystallization: the photoactive phase crystallizes from a chemically mixed reservoir, and absolute $PbBr_2$ thickness, rather than nominal $PbBr_2/PbI_2$ ratio, sets the final bandgap. In NBG stacks, by contrast, an early Pb-rich perovskite phase forms during formamidinium iodide (FAI) deposition and consolidates faster than Sn and Pb can interdiffuse, leaving persistent vertical chemical gradients. Guided by these mechanistic insights, we develop high-quality perovskite absorbers with broad bandgaps spanning 1.26–1.96 eV and demonstrate the first evaporated monolithic all-perovskite tandem in which both absorbers are formed by sTE, reaching 19.2% efficiency. Encapsulated tandems retain on average 80% of their initial efficiency after 1,200 h at 65 °C under ISOS-D-2 conditions.

## Broad bandgap tuning via sequential thermal evaporation

In our sTE route, metal-halide precursors were deposited layer-by-layer on hole transport layer (HTL)-coated indium tin oxide (ITO) substrates, followed by FAI deposition and thermal annealing to drive solid-state conversion (**Fig. 1a**). We varied the absorber bandgap by changing the inorganic precursor stack: a $PbI_2$/FAI stack provided the MBG absorber, $PbBr_2$ was introduced to supply bromide for mixed-halide WBG absorbers ($PbBr_2$/$PbI_2$/FAI stack), and $SnI_2$ was incorporated to access Sn–Pb NBG absorbers ($SnI_2$/$PbI_2$/FAI stack). For the MBG and WBG stacks, a small amount of CsI (~5 mol% relative to the B-site cations) was deposited after the FAI layer to promote structural stabilization.[17,20]

By tailoring the stack composition and annealing conditions, we developed single-junction devices with broad bandgaps spanning 1.26–1.96 eV; how different stacks convert into the final perovskite is examined in the following sections. The EQE onset shifted systematically from the Sn–Pb NBG to the mixed-halide WBG absorbers (**Fig. 1b**), and the bandgaps given here were determined from these spectra.[21] Current density–voltage (*J*–*V*) measurements confirmed their photovoltaic performance across the full bandgap range (**Fig. 1c**). The corresponding photovoltaic parameters are summarized in **Supplementary Table 1**, and the efficiency–bandgap distribution is shown in **Fig. 1d**. The NBG and WBG perovskite solar cells (PSCs) delivered power conversion efficiencies (PCEs) of 15.77% and 16.95% at 1.26 and 1.78 eV, respectively. A 1.70 eV WBG device independently measured at the National Photovoltaic Industry Measurement and Testing Center (NPVM, China) delivered a PCE of 19.8% through a 0.088-$cm^2$ aperture mask (**Supplementary Fig. 1**).

However, this bandgap tunability is not achieved simply by changing the nominal thickness ratio of the precursor layers. A $PbI_2$/FAI stack contains only one metal and one halide, so the composition of the resulting MBG absorber is fixed; as expected, its optical bandgap stayed near 1.54 eV over a broad range of thickness combinations (**Supplementary Figs. 2 and 3**). In the WBG and NBG stacks, by contrast, two chemically distinct inorganic precursors must mix during solid-state conversion, and the final absorber composition depends on how they redistribute and convert. We therefore focus on the WBG and NBG stacks to determine how initially layered precursors evolve into photoactive absorbers, and when their final compositions and bandgaps are established.

## Composition tuning in wide-bandgap stacks

To widen the absorber bandgap, we first increased the $PbBr_2$ thickness from 0 to 200 nm at a fixed $PbI_2$ thickness (220 nm) in the $PbBr_2$/$PbI_2$/FAI stack (**Fig. 2a**), raising the nominal $PbBr_2$ thickness fraction, defined as $PbBr_2$/($PbBr_2$ + $PbI_2$), from 0% to 48%. The EQE onset shifted progressively to shorter wavelengths, and the EQE-derived bandgap increased from 1.54 to 1.75 eV (**Fig. 2b**), confirming that $PbBr_2$ is an effective bromide source for widening the bandgap in WBG sTE stacks. This trend initially suggested a direct relation between the nominal $PbBr_2$ fraction and the final bandgap. However, this interpretation was challenged when the total inorganic-stack thickness was increased by scaling the $PbBr_2$/$PbI_2$ layers from 150/130 nm to 180/160 nm. Despite a slight reduction in the nominal $PbBr_2$ fraction, the EQE onset shifted to shorter wavelengths (**Fig. 2c, Supplementary Fig. 4**). Across these stacks, the bandgap increased approximately linearly with the absolute $PbBr_2$ thickness, indicating that absolute $PbBr_2$ supply, rather than nominal $PbBr_2$ fraction, determines the final bandgap of the WBG absorber.

To investigate how the fraction of $PbI_2$ affects the bandgap, we independently varied the $PbI_2$ thickness in the WBG stack. The response contrasted sharply with that observed for $PbBr_2$. Increasing the $PbI_2$ thickness from 180 to 220 nm produced a negligible shift in the EQE onset (**Fig. 2d**), even though proportional incorporation of the additional iodide would be expected to narrow the bandgap. Instead, the thicker $PbI_2$ layer mainly suppressed the short-wavelength photoresponse, suggesting that excess $PbI_2$ remained unreacted and absorbed parasitically instead of supplying iodide to the photoactive phase. X-ray diffraction (XRD) measurements supported this interpretation, showing residual crystalline $PbI_2$ in the WBG stack with a thick $PbI_2$ layer (**Supplementary Fig. 5**). These observations are inconsistent with a simple local-reaction model based on precursor proximity. Although $PbI_2$ is directly adjacent to FAI and $PbBr_2$ is separated from FAI by this intervening $PbI_2$ layer, adding $PbI_2$ does not proportionally increase the amount of iodide incorporated into the photoactive phase. A similar insensitivity held for the organic component: varying the FAI thickness induced only a modest red shift despite large absolute thickness changes (**Supplementary Fig. 6**).

The weak dependence of the bandgap on $PbI_2$ thickness prompted us to examine the limiting case in which $PbI_2$ was progressively reduced and then fully removed. Reducing the $PbI_2$ thickness from 140 nm to 0 nm changed the bandgap far less than comparable variations in $PbBr_2$ thickness (**Fig. 2b,e**), indicating that the PbI2 layer plays only a minor role in setting the bandgap of WBG stacks. Unexpectedly, even the binary $PbBr_2$/FAI stack without $PbI_2$ gave bandgaps close to 1.8 eV, well below the 1.97 eV of a spin-coated film with $FAPbBr_2I$ stoichiometry (5% Cs; **Supplementary Fig. 7**). We reasoned that this discrepancy arises mainly from the excess FAI in the evaporated stacks. Accordingly, bandgaps were plotted against a nominal Br/(Br + I) ratio that includes halides from both the inorganic precursors and FAI (**Fig. 2f** and **Supplementary Table 2**). On this basis, the spin-coated films and the ternary stacks follow the reported bandgap–composition relation,[22] whereas the binary stacks lie about 0.06 eV above it. Excess FAI therefore supplies iodide to the binary absorbers, lowering their bandgaps below the stoichiometric value, but only partly, probably because some FAI is lost during their longer annealing (see below).

Despite this decoupling between precursor stoichiometry and absorber composition, the binary and ternary stacks delivered comparable device performance in the ~1.8 eV regime (**Fig. 2g**). The difference lies instead in formation kinetics: the binary stacks reached this performance only after 120 min of annealing, whereas ternary stacks with a 10 or 20 nm $PbI_2$ interlayer already yielded functional devices after 90 min (**Fig. 2h**). Further experiments showed that inserting a thin $PbI_2$ layer between $PbBr_2$ and FAI, despite increasing the nominal precursor-stack thickness, markedly reduced the sensitivity of device performance to annealing time (**Supplementary Fig. 8**). Thus, $PbI_2$ may act as a kinetic promoter: it is not required to reach the tandem-relevant bandgap but accelerates absorber formation and widens the viable annealing window. At the same time, omitting $PbI_2$ enables a simpler binary $PbBr_2$/FAI architecture, which eliminates both an additional deposition step and the associated thickness-control variable. Unlike previously reported evaporated WBG formulations that rely on separate iodide- and bromide-containing inorganic precursors,[7,8] the binary stack achieves a ~1.8 eV absorber using $PbBr_2$ as the sole inorganic lead-halide precursor.

## Rapid halide mixing precedes slow 3D phase formation

To resolve the formation mechanism underlying this decoupled bandgap behavior, we applied a time-sliced ex situ strategy to capture the temporal evolution of precursor redistribution, intermediate-phase formation and photoactive-phase crystallization in the WBG stack. Distinct color changes during annealing defined the characteristic stages selected for analysis, allowing us to reconstruct the conversion sequence from chemically layered precursors to the final photoactive absorber.

The as-deposited $PbBr_2$/$PbI_2$/FAI stack was reddish and first turned yellow rather than darkening. After short annealing (~1 min), the stack was fully yellow, whereas a dark appearance developed only after prolonged thermal treatment (**Supplementary Fig. 9**, **Supplementary Video 1**). By contrast, the MBG $PbI_2$/FAI stack darkened within 10 s under the same annealing conditions (**Supplementary Fig. 10**, **Supplementary Video 2**) and developed both its full photoactive absorption edge and the characteristic 3D perovskite diffraction pattern on the same timescale **(Supplementary Fig. 11**). This comparison indicates that introducing $PbBr_2$ into the WBG stack shifts photoactive-phase formation from the seconds timescale of the MBG stack to tens of minutes, pointing to a distinct conversion pathway and mechanism.

To understand this conversion, we first established the initial morphological and elemental distribution of the WBG stack. Cross-sectional scanning electron microscopy (SEM) of the $PbBr_2$/$PbI_2$ template before organic-salt deposition showed a compact film with limited contrast between the two layers (**Fig. 3a**). Depth-resolved time-of-flight secondary-ion mass spectrometry (ToF-SIMS) showed that the two halides remained chemically layered, with distinct Br- and I-rich regions (**Supplementary Fig. 12**). After FAI/CsI deposition, the as-deposited full stack was approximately three times as thick as the inorganic template and showed a stratified cross-section (**Fig. 3b**). ToF-SIMS depth profiling resolved the corresponding chemical distribution: iodine and bromine remained strongly depth-dependent across the stack, whereas FA-containing species had already penetrated the inorganic precursor layers (**Supplementary Fig. 13**). Consistent with this interaction, no crystalline $PbBr_2$ reflection was detected after FAI/CsI deposition (**Supplementary Fig. 14**), showing that the $PbBr_2$ lattice had already been disrupted before annealing.

During early-stage annealing, precursor redistribution and intermediate-phase formation outpaced photoactive-phase formation. Within 1 min, cross-sectional SEM showed that the upper organic-rich region had merged with the underlying inorganic template, largely erasing the initially layered morphology (**Fig. 3b and Supplementary Figs. 15 and 16**). This rapid reorganization produced relatively large, perovskite-like grain domains. These domains were transient, reconstructing into smaller grains at intermediate annealing times before coarsening again during long-term annealing. The $Br_2^-$ depth profile lost most of its stratification between 8 and 16 s and became uniform by 60 s **(Fig. 3c**; the $I_2^-$ profiles are shown in **Supplementary Fig. 17**). The crystallographic evolution of the photoactive perovskite occurred on a markedly longer timescale. The higher-angle feature near 14.78° in the as-deposited stack disappeared within 8 s, whereas the feature near 14.3° shifted to higher 2θ and gained intensity mainly between 1 and 60 min (**Fig. 3d and Supplementary Fig. 18**). We attribute the former to a Br-rich phase, since it lies close to the (100) reflection of FAPbBr3. The shift of the latter to higher 2θ reflects lattice contraction as the smaller $Br^-$ replaces $I^-$, as expected from Vegard's law for mixed I–Br perovskites.[17] This points to the gradual formation of a mixed I–Br perovskite phase during prolonged annealing. Consistently, no well-defined photoactive absorption edge was observed through 15 min; the edge appeared at 30 min and sharpened further between 60 and 120 min (**Supplementary Fig. 19**).

The same XRD series also revealed an intermediate phase underlying the delayed conversion into the photoactive perovskite phase. The as-deposited stack exhibited sharp features near 24.76° and 25.63° (**Fig. 3e**; raw data in **Supplementary Fig. 20**), both of which have been reported for crystalline FAI.[12] Upon annealing, the sharp doublet collapsed within 8 s, and was replaced by a broad feature near 25°, indicating that crystalline FAI is disrupted almost immediately. We also resolved three additional low-angle reflections at 6.81°, 10.04° and 10.73°, which appeared within the first 8 s of annealing and vanished by 60 min (**Supplementary Fig. 21**). These reflections provide diffraction evidence that the precursors reorganize into a transient intermediate phase, and their disappearance as the 3D perovskite reflection grows links this intermediate phase to subsequent photoactive-phase formation.

Together, these observations resolve the conversion pathway of the WBG stack into the six stages illustrated in **Fig. 3f**. After FAI/CsI deposition, FA-containing species readily penetrate the $PbBr_2/PbI_2$ template, so FA is already distributed throughout the template before annealing. The resulting as-deposited stack is substantially expanded and contains only a minor early perovskite phase together with excess crystalline FAI, while Br and I remain vertically segregated. During early-stage annealing, crystalline FAI collapses and transient intermediate reflections emerge together with rapid Br and I redistribution, yielding a reservoir with homogenized halides that remains iodide-rich. The 3D perovskite phase forms from this reservoir during prolonged annealing, while excess FAI is volatilized and the film densifies (**Fig. 3b**). This temporal separation indicates that long-range halide interdiffusion is not the kinetic bottleneck; instead, WBG conversion is limited by reconstruction of the reservoir into the photoactive 3D phase. The final bandgap is therefore set during late crystallization from this reservoir, and it tracks the $PbBr_2$ supply, the only bromide source in the stack. This reservoir resembles a precursor solution in providing chemically mixed species for crystallization, which explains why excess FAI-derived iodide can lower the bandgap of binary stacks, whereas the volatilization of excess FAI during prolonged annealing (**Fig. 3b**) accounts for their deviation from the empirical relation (**Fig. 2f**).

## Slow cation interdiffusion in NBG stacks

Having established the conversion pathway of the WBG perovskite, we next applied the same time-sliced ex situ strategy to the Sn–Pb NBG stack. NBG absorbers require balanced Sn and Pb incorporation on the B site.[18,23] We therefore introduced a $SnI_2$ layer beneath $PbI_2$ in the $PbI_2$/FAI stack. This modification produced a conversion process that contrasts sharply with that of the WBG stacks. Whereas the WBG stack remained yellow during early annealing, the NBG stack darkened within seconds, indicating much faster formation of an optically active perovskite-related phase (**Supplementary Fig. 22** and **Supplementary Video 3**).

The as-deposited $SnI_2$/$PbI_2$ inorganic template appeared morphologically compact in cross-section (**Fig. 4a**), but retained clear chemical stratification, with separate Sn-rich and Pb-rich regions resolved by ToF-SIMS (**Supplementary Fig. 23**). Deposition of FAI increased the stack to approximately twice the thickness of the inorganic template and reorganized it into a pronounced three-layer structure comprising an FA-rich upper region, a Pb-rich middle region and a Sn-rich lower region (**Fig. 4b and Supplementary Fig. 24**). Unexpectedly, the Sn-rich region became porous and contained a voided zone adjacent to the substrate, as independently confirmed by high-angle annular dark-field scanning transmission electron microscopy (HAADF-STEM; **Supplementary Fig. 25**). During the first seconds of annealing, fine perovskite-like grains appeared at the surface, while the internal layer boundaries became indistinguishable and the voids started to merge. Further annealing promoted grain coalescence and the development of columnar grains; the buried voids merged and were no longer resolved after prolonged annealing.

XRD showed that an early perovskite phase forms during FAI deposition and continues to reorganize during annealing. A strong perovskite phase reflection near 14° was already present in the as-deposited stack; the reaction had therefore begun during FAI deposition and before thermal annealing (**Fig. 4c**). By contrast, the $PbI_2$/FAI MBG reference developed its characteristic 3D perovskite diffraction signature only after annealing began, within approximately 10 s (**Supplementary Fig. 11**). This comparison indicates that the buried $SnI_2$ layer markedly accelerates perovskite formation, shifting its onset from thermal annealing to the deposition stage. During early-stage annealing, the NBG perovskite reflection shifted to higher $2\theta$ as the $SnI_2$- and $PbI_2$-related diffraction features rapidly diminished, consistent with progressive Sn incorporation into an initially Pb-rich lattice (**Supplementary Fig. 26**). Yet the integrated area of the perovskite reflection decreased to approximately half of its initial value after 1 min (**Fig. 4c and Supplementary Fig. 27**). This non-monotonic evolution is inconsistent with continuous crystal growth and instead indicates substantial lattice reorganization during precursor consumption.

The chemical profiles locate this early reaction near the original $SnI_2$/$PbI_2$ boundary. In the as-deposited $SnI_2$/$PbI_2$/FAI stack, the $CH_5N_2^+$ signal penetrated through the Pb-rich region but decayed within the upper part of the Sn-rich region (**Fig. 4d**). FA-containing species therefore penetrated the NBG inorganic template less completely than in the WBG stack, where penetration was already complete before annealing (**Supplementary Fig. 13**). The coincidence of this truncated FA profile with the lower $2\theta$ perovskite reflection in the as-deposited stack than that of the annealed film supports a reaction-front mechanism. We therefore propose that FA-containing species pass through the Pb-rich region, forming an initial Pb-rich perovskite phase near the $SnI_2/PbI_2$ interface. The reacted phase could then restrict further FA penetration, leaving the lower Sn-rich region comparatively FA-poor. In this picture, Sn-containing species redistribute upward from the buried Sn-rich region to supply the growing perovskite phase, while further downward FA penetration is restricted by the reacted phase. If this outward transport is not compensated by inward material transport, local depletion could contribute to the transient buried porosity, analogous to Kirkendall-type voiding arising from unbalanced fluxes across a reactive interface.[24] Its subsequent disappearance is consistent with continued mass redistribution during annealing.

To quantify Sn/Pb interdiffusion, we calculated the relative Sn-containing secondary-ion fraction ( $f_{\mathrm{Sn}} = \frac{I_{\mathrm{Sn}I_2^-}}{I_{\mathrm{SnI}_2^-}+I_{\mathrm{PbI}_2^-}}$ ) near the surface and the buried interface (**Fig. 4e and Supplementary Fig. 28**). At the buried interface, $f_{\mathrm{Sn}}$ decreased from approximately 100% in the as-deposited stack to 56% after 60 min, whereas the corresponding value near the surface increased from approximately 0 to only 35%. These opposing changes demonstrate net Sn/Pb interdiffusion, while the remaining separation shows that through-thickness homogenization remains incomplete. We attribute this incomplete homogenization to rapid lattice formation and consolidation: once a solid perovskite phase has formed, further Sn/Pb mixing can no longer proceed freely through an unconverted precursor reservoir but requires transport through an established perovskite lattice. The optical evolution provides an independent signature of this slow compositional homogenization. The absorption onset shifted rapidly to an apparent bandgap below 1.5 eV between 2 and 10 s of annealing, marking the formation of optically active perovskite domains (**Supplementary Fig. 29**). This red-shift continued to ~1.26 eV after the dominant perovskite diffraction-peak position had nearly stabilized. Local Sn/Pb mixing therefore continues after the average lattice parameter has stabilized. Because $Sn^{2+}$ is only slightly smaller than $Pb^{2+}$, Sn- and Pb-based iodide perovskites have similar lattice parameters,[25] so the diffraction-peak position is relatively insensitive to the Sn/Pb distribution. The bandgap, in contrast, depends strongly on the local Sn/Pb arrangement.[18,26]

These results support the NBG conversion pathway illustrated in **Fig. 4f**. Perovskite formation begins during FAI deposition, before either FA or the metal cations are homogenized. As FA reaches the $SnI_2/PbI_2$ boundary, a localized Pb-rich perovskite phase forms and restricts further FA penetration into the buried Sn-rich region. Early-stage annealing partially overcomes this transport barrier, consuming the remaining inorganic precursors and initiating Sn/Pb exchange as the layered morphology collapses and the buried voids coalesce. The perovskite lattice, however, reorganizes and consolidates faster than the metal cations can homogenize. The same rapid phase formation that creates the photoactive absorber thus progressively constrains the cation exchange required for B-site homogenization. Prolonged annealing coarsens the grains and closes the buried pores, yet Sn/Pb redistribution remains incomplete, and a pronounced vertical chemical gradient persists. NBG stack conversion in sTE is therefore limited not by photoactive-phase formation but by post-formation equilibration of the B-site cations.[27] The persistent gradient leaves substantial scope for further improvement by increasing Sn/Pb interdiffusion. This contrasts strongly with the WBG conversion pathway and highlights the need for bandgap-specific kinetic control during annealing.

## Monolithic all-perovskite tandems by sequential evaporation

The divergent conversion mechanisms identified above guided the bandgap-specific optimization of the WBG and NBG absorbers. On this basis, we demonstrate, to our best knowledge, the first monolithic all-perovskite TSC in which both WBG and NBG absorbers are formed by sTE.

**Figure 5a** shows the p–i–n monolithic tandem architecture. To verify the integrity of the multilayer stack, we performed cross-sectional scanning transmission electron microscopy–energy-dispersive X-ray spectroscopy (STEM–EDX) mapping and elemental line-profile analysis, which confirmed the intended layer sequence and the spatial separation of the two absorbers (**Fig. 5b**). The best-performing tandem device delivered a PCE of 19.2%, with an open-circuit voltage ($V_{OC}$) of 1.85 V, a short-circuit current density ($J_{SC}$) of 13.12 mA $cm^{-2}$ and a fill factor (FF) of 79.44% (**Fig. 5c**), together with a stabilized power output (SPO) of 18.5% (**Supplementary Fig. 30**). EQE spectra confirmed photoresponse from both the WBG and NBG subcells (**Fig. 5d**). Continuous maximum-power-point tracking (MPPT) under 100 mW $cm^{-2}$ illumination further showed a $T_{80}$ lifetime exceeding 130 h (**Fig. 5e**). Under ISOS-D-2 storage at 65 °C,[28] encapsulated tandem devices retained, on average, 80% of their initial efficiency after 1,200 h (**Fig. 5f**).

## Conclusions

Using a time-sliced ex situ strategy, we have shown that WBG and NBG stacks in sTE convert into photoactive absorbers through divergent solid-state pathways, each governed by distinct kinetic bottlenecks. In WBG stacks, conversion is limited by slow reconstruction of a chemically mixed reservoir into the photoactive phase, which makes $PbBr_2$ supply the primary determinant of the final bandgap; in NBG stacks, by contrast, early Pb-rich perovskite formation and rapid lattice consolidation outpace Sn/Pb interdiffusion, leaving vertical B-site gradients. Applying this understanding, we developed perovskite absorbers with bandgaps spanning 1.26–1.96 eV and demonstrated, to our best knowledge, the first evaporated all-perovskite tandem in which both absorbers are formed by sTE, reaching a PCE of 19.2%. Furthermore, encapsulated tandems retained on average 80% of their initial efficiency after 1,200 h at 65 °C under ISOS-D-2 conditions. This work establishes bandgap-specific control of solid-state conversion as a design principle for sequentially evaporated perovskite tandem photovoltaics. Narrowing the remaining efficiency gap to solution-processed tandems will require accelerating photoactive-phase formation in WBG stacks, and improving Sn/Pb interdiffusion in NBG absorbers through tailored annealing protocols.

## Methods

### Materials

Pre-patterned indium tin oxide (ITO) glass substrates (15 Ω $sq^{-1}$) were purchased from Advanced Election Technology Co., Ltd. Lead(II) iodide ($PbI_2$, 99.999%), lead(II) bromide ($PbBr_2$, 99.999%), tin(II) iodide ($SnI_2$, 99.999%) and Hellmanex III were obtained from Sigma-Aldrich. Copper (Cu, ≥99.99%) was purchased from Angstrom Engineering. Formamidinium iodide (FAI, >99.99%) was acquired from Greatcell Solar Ltd. Fullerene-C60 and bathocuproine (BCP) were purchased from Xi'an Yuri Solar Co., Ltd. 4PADCB was obtained from Luminescence Technology Corp. (Lumtec). Caesium iodide (CsI) was acquired from Tokyo Chemical Industry Co., Ltd. (TCI). Ethanol (analytical grade) was obtained from VWR. Poly(3,4-ethylenedioxythiophene):polystyrene sulfonate (PEDOT:PSS, Clevios P VP AI 4083) was purchased from Heraeus, LLC. All chemicals used in this work were commercially available and used as received.

## Film and device fabrication

**Films and single-junction devices.** Pre-patterned ITO substrates were first soaked in a 2% aqueous Hellmanex III solution for 60 min, followed by sequential sonication in deionized water and isopropanol for 10 min each. The substrates were then dried under a nitrogen flow and treated with UV–ozone for 30 min to improve surface cleanliness and wettability.

For WBG and MBG devices, 4PADCB was used as the hole-selective self-assembled monolayer. The 4PADCB solution was prepared in ethanol at a concentration of 0.3 mg $ml^{-1}$. The solution was dispensed onto the cleaned ITO substrates and allowed to rest for 30 s, followed by spin coating at 3,000 rpm for 30 s and thermal annealing at 100 °C for 2 min. For NBG devices, PEDOT:PSS (Clevios P VP AI 4083) was filtered through a 0.45 μm hydrophilic polyvinylidene fluoride (PVDF) filter and spin-coated onto cleaned ITO substrates in two steps: 1,000 rpm for 1 s (acceleration 1,000 rpm $s^{-1}$), then 4,000 rpm for 40 s (acceleration 2,000 rpm $s^{-1}$). The films were then annealed at 140 °C for 20 min.

The HTL-coated substrates were subsequently transferred into a vacuum chamber for sequential thermal evaporation of the perovskite precursor layers. FAI, $SnI_2$, $PbI_2$ and $PbBr_2$ were deposited at a rate of 1 Å $s^{-1}$, while CsI was deposited at 0.15 Å $s^{-1}$. After precursor deposition, WBG and MBG films were removed from the vacuum chamber and annealed in ambient air at 170 °C for 8 s to 120 min under a relative humidity of 30–40% to complete perovskite formation. NBG perovskite films were annealed in a $N_2$-filled glovebox at 150 °C for 2 s to 60 min.

After annealing, the samples were transferred back into the vacuum chamber for charge-transport layer and electrode deposition. $C_{60}$ (23 nm), BCP (7 nm) and Cu (100 nm) were sequentially deposited under a base pressure of ~$1 \times 10^{-6}$ mbar. The active area of the devices was defined by a patterned Cu shadow mask.

**Tandem devices.** Monolithic all-perovskite tandem devices were fabricated following the WBG single-junction process up to the deposition of $C_{60}$. The BCP layer used in single-junction devices was replaced by a 20 nm $SnO_x$ deposited by atomic layer deposition at 100 °C. A 1.5 nm Au layer was then thermally evaporated onto the $SnO_x$ layer to form the interconnecting contact. The NBG bottom subcell was subsequently fabricated on top of the interconnecting layer. PEDOT:PSS was diluted with methanol at a volume ratio of 1:2, spin-coated using the same parameters as for NBG single-junction devices and annealed at 100 °C for 10 min. The remaining steps for NBG absorber formation and top-contact deposition were identical to those used for the NBG single-junction devices. Both the NBG devices and tandem devices were encapsulated with UV-curable epoxy and a cover glass for characterization outside the $N_2$-filled glovebox.

**Calculation of Br ratio.** Nominal Br/(Br + I) ratios were obtained by converting the deposited layer thicknesses into molar amounts using the densities of $PbBr_2$ (6.66 g cm$^{-3}$), $PbI_2$ (6.16 g cm$^{-3}$), CsI (4.51 g cm$^{-3}$) and FAI (2.48 g cm$^{-3}$), counting two halide ions per $PbX_2$ and one per FAI or CsI.[29,30]

## Characterization

***J–V* and EQE measurements**. Current density–voltage (*J–V)* characteristics were measured in four-contact mode at standard test conditions (100 mW cm$^{-2}$) using a Keithley 2400 source meter. A solar simulator (ABA class, LOT-QuantumDesign) was calibrated to AM 1.5G one sun illumination using a certified monocrystalline silicon solar cell (RS-ID-5, Fraunhofer-ISE). No spectral-mismatch correction was applied. The *J–V* measurements were performed in both reverse and forward directions at a scan rate of 100 mV s$^{-1}$ (voltage step, 10 mV), with the devices held at 25 °C. The steady-state efficiency as a function of time was recorded using a maximum power point (MPP) tracker, which adjusts the applied voltage to reach the maximum power point (perturb and observe algorithm). The external quantum efficiency of the solar cells was measured with a lock-in amplifier. The probing beam was generated by a chopped white source (900 W, halogen lamp, 269 Hz) and a dual grating monochromator. The beam size was adjusted to ensure an illumination area within the cell area. A certified single crystalline silicon solar cell was used as a reference cell. White bias light was applied during the measurement with an intensity of ~0.1 sun. For tandem EQE measurements, extra light/voltage bias was applied to selectively measure the WBG and NBG subcells. The WBG subcell was measured under halogen bias light saturating the NBG subcell with 0.6 V applied to the device, and the NBG subcell under LED bias light saturating the WBG subcell with 1.0 V applied; the voltage bias compensates the photovoltage of the biased subcell so that the measured subcell is held near short-circuit. All *J–V* and EQE measurements were made through a black metal aperture mask with an opening of 0.0616 cm$^2$ (NBG single junctions and tandems) or 0.088 cm$^2$ (MBG and WBG single junctions); the aperture openings were measured by optical microscopy. Efficiencies are reported for the aperture area.

**SEM and XRD.** Scanning electron microscopy (SEM) images were obtained using a Hitachi S-4800 scanning electron microscope operated at an acceleration voltage of 5 kV. X-ray diffraction (XRD) patterns were recorded using an X'Pert Pro diffractometer (PANalytical) in Bragg–Brentano geometry with Cu $K\alpha_1$ radiation ($\lambda$ = 1.5406 Å). The diffraction data were collected over a 2θ range of 5° to 60°, with a step size of 0.0167°. All patterns were aligned to the $In_2O_3$ (400) reflection of the ITO substrate as an internal standard.

**ToF-SIMS.** Time-of-flight secondary ion mass spectrometry (ToF-SIMS) measurements were performed on an IONTOF V system. The primary beam was 25 keV $Bi^{3+}$ with a raster size of 50 × 50 μm$^2$. $Cs^+$ ions were used with 1000 eV ion energy on a 300 × 300 μm$^2$ raster size to bombard and etch the film. The collected secondary ion spectra were normalized to the total counts. Positive and/or negative secondary ions were collected depending on the target fragments.

**Focused ion beam (FIB) STEM-EDX**. For as-deposited NBG stack, FIB-STEM were performed using a JEM-Grand ARM300CF microscope equipped with double spherical-aberration correctors and operated at an accelerating voltage of 300 kV. For the tandem device, FIB-STEM-EDX was performed on a TFS Helios 5 CX instrument. A 100 nm protective carbon layer was first deposited on the copper surface using a 2 kV, 5.5 nA electron beam. Using AutoTEM 5.9, a further 500 nm tungsten protective layer was then coated using the gallium ion beam. Rough milling of the lamella, needle trench and cutout was performed with a 30 kV, 2.5 nA ion beam. The lamella was lifted out and welded to a copper lift-out grid for thinning without breaking vacuum. Lamella thinning was performed manually to 200 nm using 1° over-/undertilt with a 30 kV, 24 pA ion beam to minimize beam damage, followed by a final thinning step to 100 nm with a 5 kV, 4 pA ion beam. STEM-EDX acquisition was performed using Aztec nanoanalysis software at 30 kV and 86 pA to minimize detector dead-time and electron-beam-induced sample degradation. The EDX map was acquired in a single pass with ~2 nm pixel width and 100 μs dwell time per pixel. The line-scan data in Fig. 5b were extracted by integrating the mapped element intensity over a 1 μm subsection of the scan window to improve the signal-to-noise ratio, and subsequently normalized for each element to improve visual clarity.

**UV–vis absorption spectroscopy.** UV–vis transmittance and reflectance spectra of the perovskite films were measured using a Shimadzu UV/Vis 3600 spectrophotometer equipped with an integrating sphere. Reflectance spectra were corrected for the instrumental response arising from diffuse and specular reflections from the sample. Absorption spectra were calculated from the measured transmittance and corrected reflectance spectra, and optical bandgaps were extracted from Tauc plots.

**MPP tracking.** Long-term operational stability was measured with a Litos Lite system (FLUXiM AG). An encapsulated tandem device was held at its maximum power point by a perturb-and-observe algorithm under continuous illumination from a white-LED solar simulator (intensity equivalent to 100 mW $cm^{-2}$) in a continuous $N_2$ flow. The device temperature, set by the illumination without active heating or cooling, was 54–57 °C. Voltage and current at the maximum power point were recorded every 5 min; no preconditioning was applied. These conditions correspond to the ISOS-L-2I procedure except that the device temperature was 54–57 °C rather than 65 °C, and the test is therefore reported with its full conditions.

**ISOS-D-2 testing.** Eight tandem devices encapsulated with UV-curable epoxy and a cover glass were stored in a dark oven at 65 °C in ambient air (uncontrolled humidity) without preconditioning. Devices were removed periodically, cooled to room temperature and measured by J–V under the same simulator and aperture mask as for the initial characterization. Two devices lost their external Ag contacts during the test and were excluded; the remaining six are shown in **Fig. 5f**, each normalized to its own initial efficiency, together with the mean ± s.d.

# Data availability

All data supporting the findings of this study are available within the Article and its Supplementary Information. Any additional data requests can be directed to the corresponding authors. Source data are provided with this paper.

## Acknowledgements

This work has received funding from the European Union's Horizon Europe research and innovation programme under grant agreement No. 101075605, No. 101147311 and the Horizon 2020 research and innovation programme under grant agreement No. 851676. This work has also been supported by the Swiss National Science Foundation (grant no. 200021_213073) and the Swiss Federal Office of Energy (SFOE, grant no. SI/502549-01). H.L. thanks the China Scholarship Council (CSC) for funding from the Ministry of Education of P. R. China. The authors gratefully acknowledge ScopeM for support and assistance in this work.

## Author contributions

F.F. and H.L. conceived the idea and F.F. directed the overall research. H.L. performed most of the sample preparation, including NBG and WBG PSCs, tandem solar cells and corresponding thin films, as well as most of the characterizations and data analysis. N.H. assisted with the preparation and optimization of WBG perovskite films and solar cells. A.W. contributed to NBG PSC and TSC fabrication and optimization as well as its ISOS-D-2 testing. J.L. performed XRD measurements and MPP tracking measurements for TSCs. N.U. contributed to WBG PSC fabrication for TSCs. F.D.G. contributed to the optimization of MBG and WBG PSCs. T.S., S.B.S. and C.S. performed STEM characterization and analysis. H.L. and F.F. wrote the original manuscript, and all authors contributed to reviewing and editing the manuscript. W.T. and F.F. supervised the project.

## Corresponding authors

Correspondence to Fan Fu (fan.fu@empa.ch); Wolfgang Tress (trew@zhaw.ch)

## Competing interests

The authors declare no competing interests.

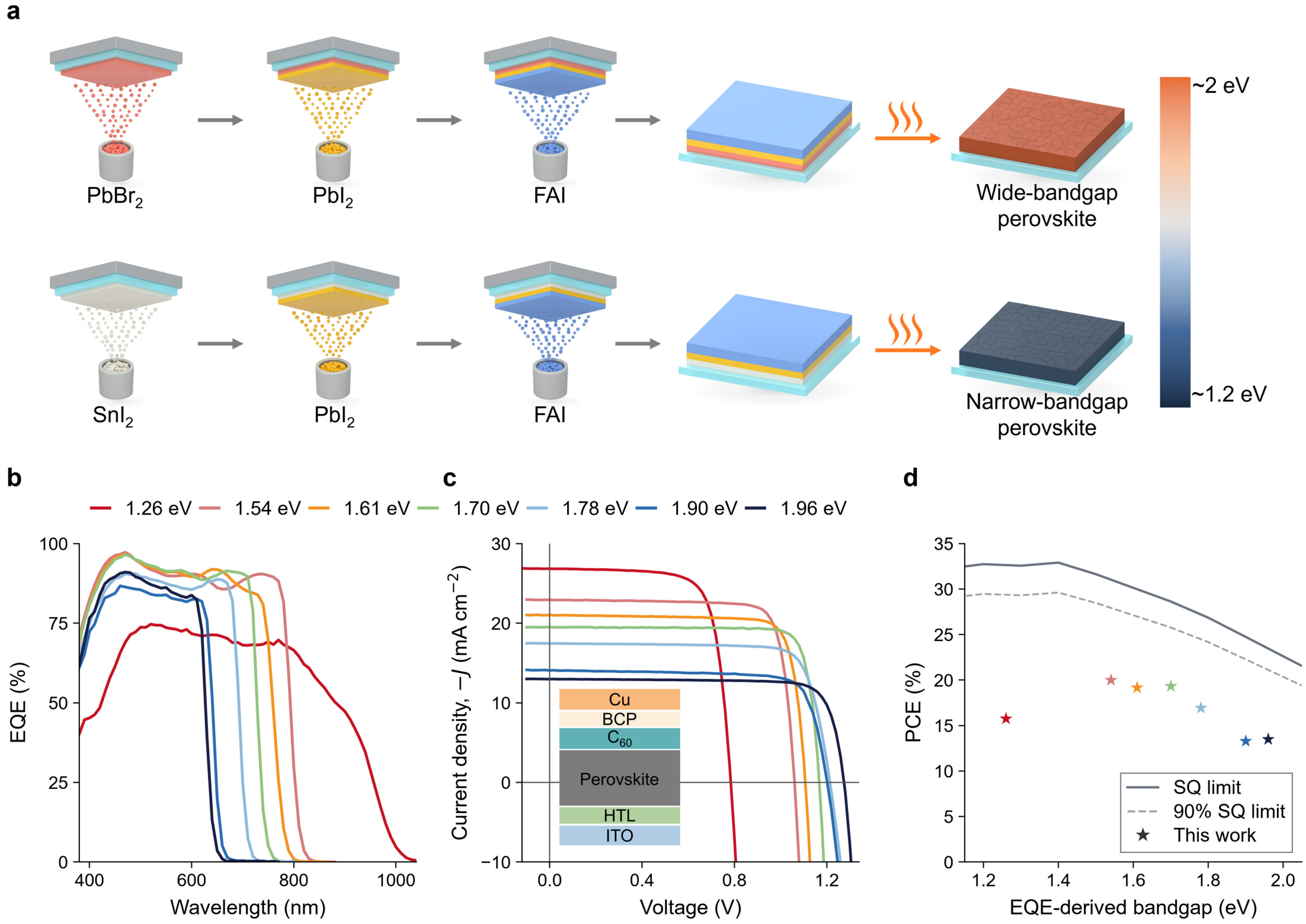

**Fig. 1 Broad bandgap tunability of perovskites by sequential thermal evaporation.** **a,** Schematic of the sTE route. **b,** External quantum efficiency (EQE) spectra of representative single-junction devices with different bandgaps. **c,** Current density–voltage (*J*–*V*) characteristics of devices with different bandgaps. Inset, device architecture used for single junction devices. **d,** Power conversion efficiency (PCE) plotted as a function of EQE-derived bandgap. Bandgaps were determined from the EQE spectra as the inflection point of the absorption edge, i.e. the maximum of $d_{EQE}/d_E$, using the open-source software grapa.[21] Solid and dashed grey lines denote the Shockley–Queisser (SQ) limit and 90% of the SQ limit, respectively; symbols denote the champion PCE measured in this work at each bandgap, with the corresponding device parameters listed in **Supplementary Table 1**.

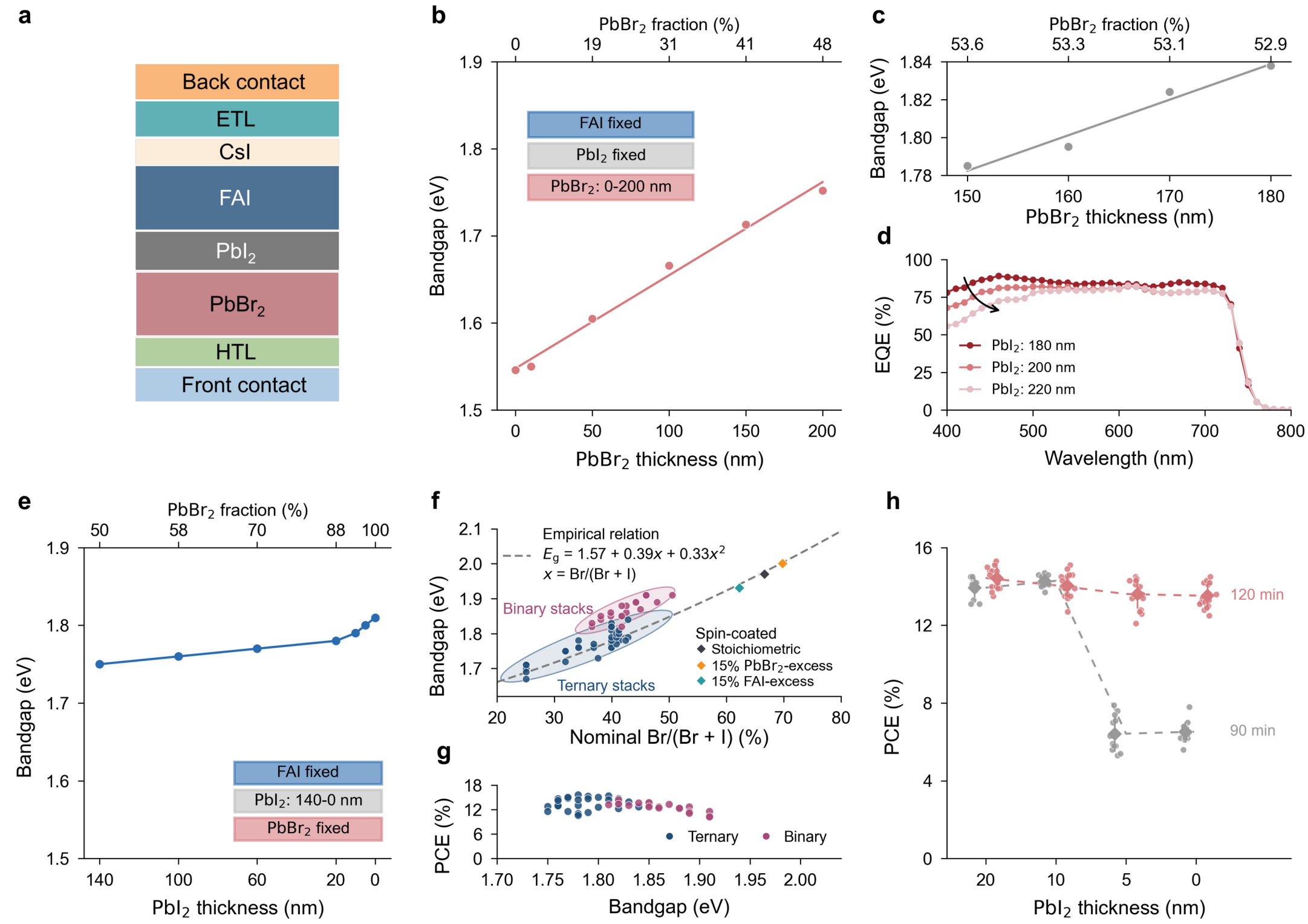

**Fig. 2 Asymmetric precursor control governs wide-bandgap perovskite formation in sTE.** **a,** Schematic of the WBG device stack, with the absorber region expanded to show the as-deposited precursor layers before conversion. **b,** EQE-derived bandgap as a function of $PbBr_2$ thickness at fixed $PbI_2$ and FAI thicknesses. **c,** EQE-derived bandgap plotted against $PbBr_2$ thickness and nominal $PbBr_2$ fraction in $PbBr_2/PbI_2$ precursor stacks. **d,** EQE spectra of devices prepared with different $PbI_2$ thicknesses at fixed $PbBr_2$ and FAI thicknesses. **e,** EQE-derived bandgap as a function of $PbI_2$ thickness and nominal $PbBr_2$ fraction. **f,** EQE-derived bandgap versus the nominal Br/(Br + I) ratio of ternary and binary stacks (layer compositions in **Supplementary Table 2**). The nominal ratio counts all halide sources and was calculated from the deposited layer thicknesses (Methods). Diamonds show spin-coated films prepared from stoichiometric, 15% $PbBr_2$-excess and 15% FAI-excess solutions, plotted at the Br/(Br + I) ratio of the solution; their bandgaps were extracted from Tauc plots. The dashed line shows the empirical relation $E_g = 1.57 + 0.39x + 0.33x^2$, with $x$ = Br/(Br + I).[22] **g,** PCE of all individual devices (n = 80: 56 from ternary and 24 from binary stacks) plotted against EQE-derived bandgap. **h,** PCE distributions (n = 16 devices per condition). Small symbols represent individual devices; diamonds and error bars denote the mean ± s.d.; dashed lines are guides to the eye.

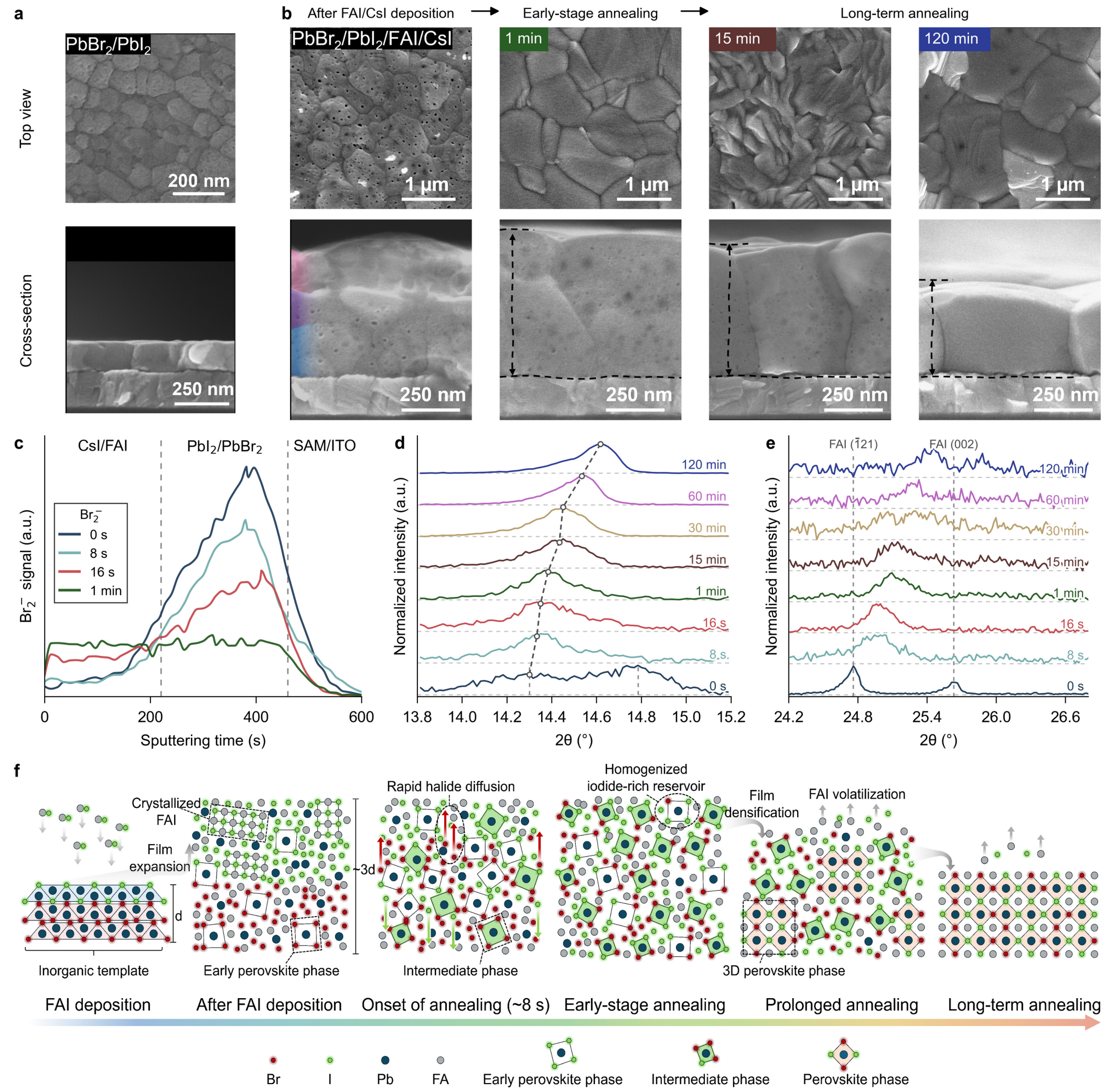

**Fig. 3 Wide-bandgap stacks undergo rapid halide redistribution but slow photoactive-phase formation.**

**a,** Top-view and cross-sectional SEM images of the as-deposited inorganic WBG template before FAI deposition. **b,** Top-view and cross-sectional SEM images of the ITO/4PADCB/$PbBr_2$/$PbI_2$/FAI/CsI stacks after different annealing times. **c,** ToF-SIMS $Br_2^-$ depth profiles during the first minute of annealing. **d,** Ex situ XRD patterns of the perovskite diffraction region after different annealing times. Dashed vertical guides mark the two perovskite-related features present in the as-deposited stack (14.3 and 14.78°); the dashed line with open markers traces the position of the perovskite peak maximum across the series. **e,** Ex situ XRD patterns of the FAI diffraction region after different annealing times. Dashed vertical lines mark the ($\bar{1}$21) and (002) reflections of monoclinic FAI. **f,** Schematic of the conversion pathway of the WBG stack, from FAI deposition on the inorganic template to the dense mixed-halide 3D perovskite. CsI is omitted for clarity.

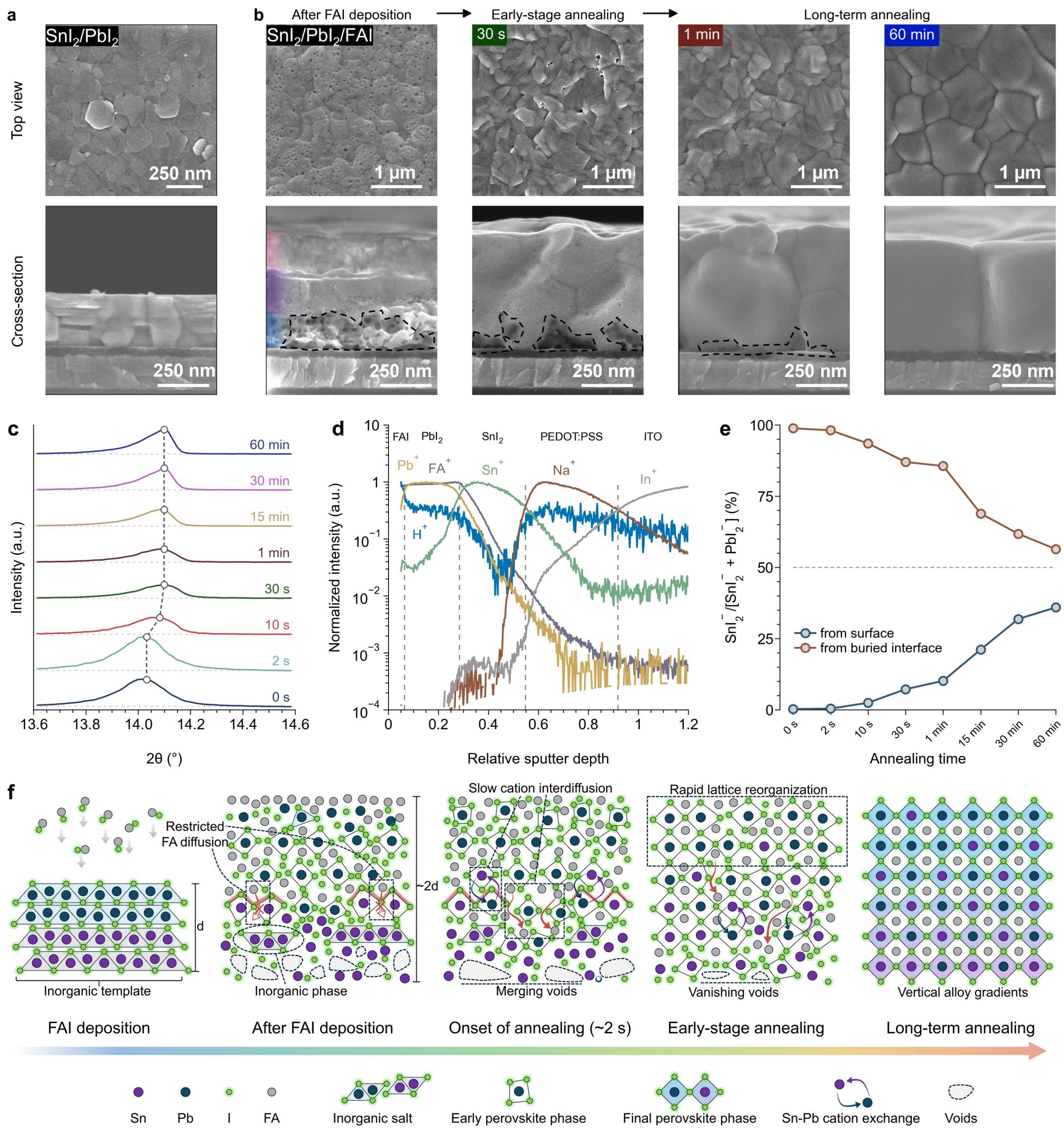

**Fig. 4 Narrow-bandgap stacks undergo rapid perovskite formation but slow cation interdiffusion.**

**a,** Top-view and cross-sectional SEM images of the as-deposited inorganic NBG template before FAI deposition. **b,** Top-view and cross-sectional SEM images of the ITO/PEDOT:PSS/$SnI_2$/$PbI_2$/FAI stacks after different annealing times. **c,** Ex situ XRD patterns of the perovskite diffraction region after different annealing times. The dashed line with open markers traces the position of the perovskite peak maximum across the series. **d,** ToF-SIMS depth profiles of the $CH_5N_2^+$, $Pb^+$, $Sn^+$, $H^+$, $Na^+$ and $In^+$ signals in the as-deposited ITO/$SnI_2$/$PbI_2$/FAI stack. Dashed vertical lines indicate the nominal FAI, $PbI_2$, $SnI_2$, PEDOT:PSS and ITO regions. **e,** Evolution of the $SnI_2^-$ signal fraction at positions 20% from the film surface and 20% from the buried interface. **f,** Schematic of the NBG conversion pathway, from FAI deposition on the inorganic template to the dense perovskite film.

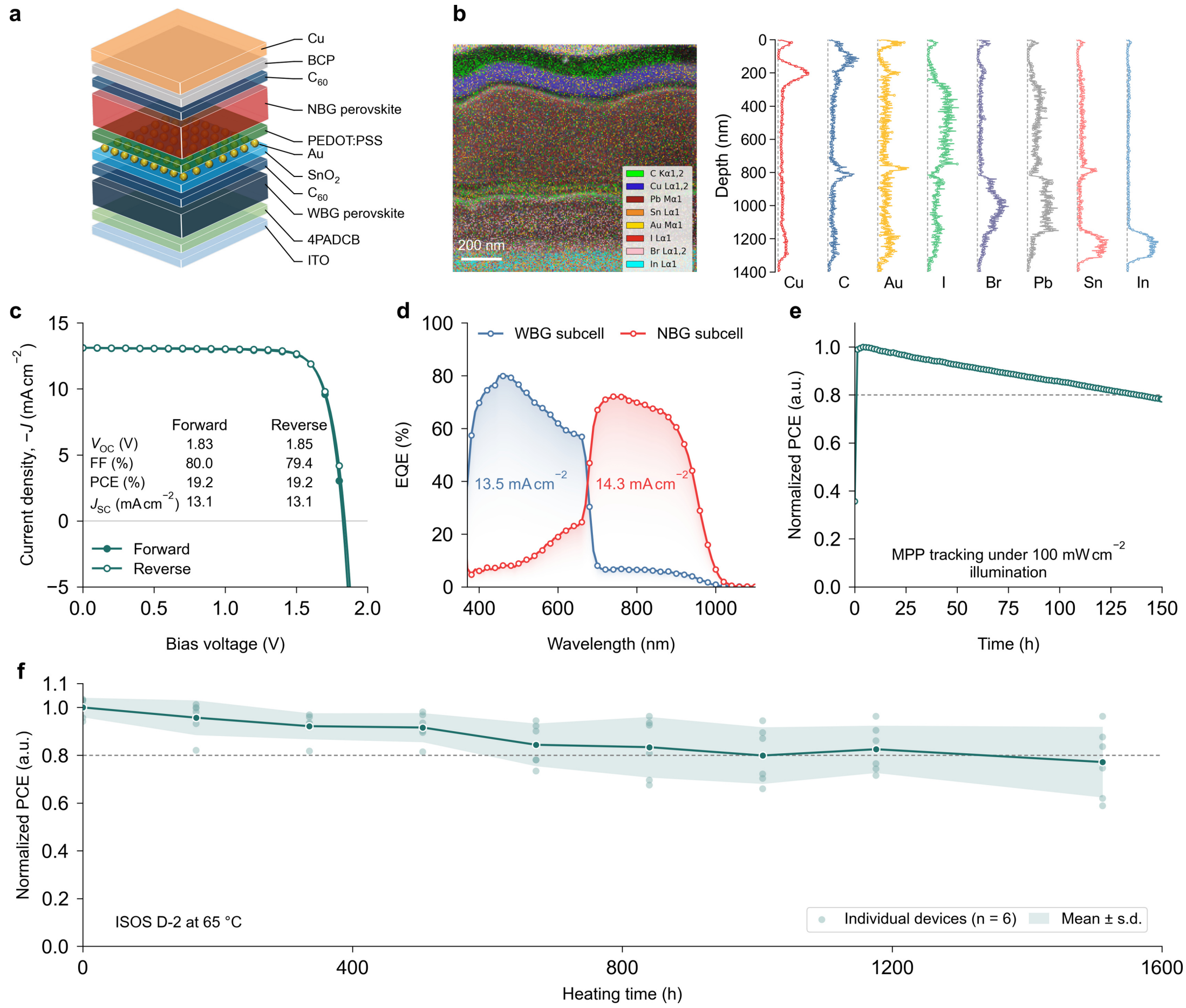

**Fig. 5 Monolithic all-perovskite tandem solar cells enabled by sTE.**

**a,** Schematic architecture of the monolithic all-perovskite tandem device. **b,** Cross-sectional STEM–EDX elemental mapping and corresponding line scan of the tandem device. **c,** Current density–voltage (*J*–*V*) characteristics of the best-performing tandem device. **d,** EQE spectra of the WBG and NBG subcells measured under subcell-selective bias light and voltage (Methods); the values give the $J_{SC}$ obtained by integrating each spectrum over the AM 1.5G spectrum. **e,** Continuous maximum-power-point tracking of an encapsulated tandem device under 1 sun illumination in continuous $N_2$ gas flow at 54–57 °C. The dashed line marks 80% of the initial efficiency. **f**, ISOS-D-2 ageing of encapsulated tandem devices stored in the dark at 65 °C (6 devices).

*Supplementary Information*

**Divergent Solid-state Conversion Pathways in Evaporated All-perovskite Tandem Solar Cells**

Huagui Lai[1,2], Amber Wright[1], Nick Huber[1], Niels Uythoven[1], Federico De Giorgi[1], Jincheng Luo[1], Tristan Sachsenweger[2], Sunil B. Shivarudraiah[3], Chih-Jen Shih[3], Wolfgang Tress[2,*], Fan Fu[1,*]

[1]Laboratory for Thin Films and Photovoltaics, Empa – Swiss Federal Laboratories for Materials Science and Technology, Dübendorf 8600, Switzerland

[2]Institute of Computational Physics, Zurich University of Applied Sciences, Winterthur 8400, Switzerland

[3]Institute for Chemical and Bioengineering, ETH Zürich, Zürich 8093, Switzerland

Correspondence and requests for materials should be addressed to F.F. (fan.fu@empa.ch) or W.T. (trew@zhaw.ch).

This file contains: Supplementary Figs. 1–30, Supplementary Tables 1 and 2, legends for Supplementary Videos 1–3 and Supplementary References.

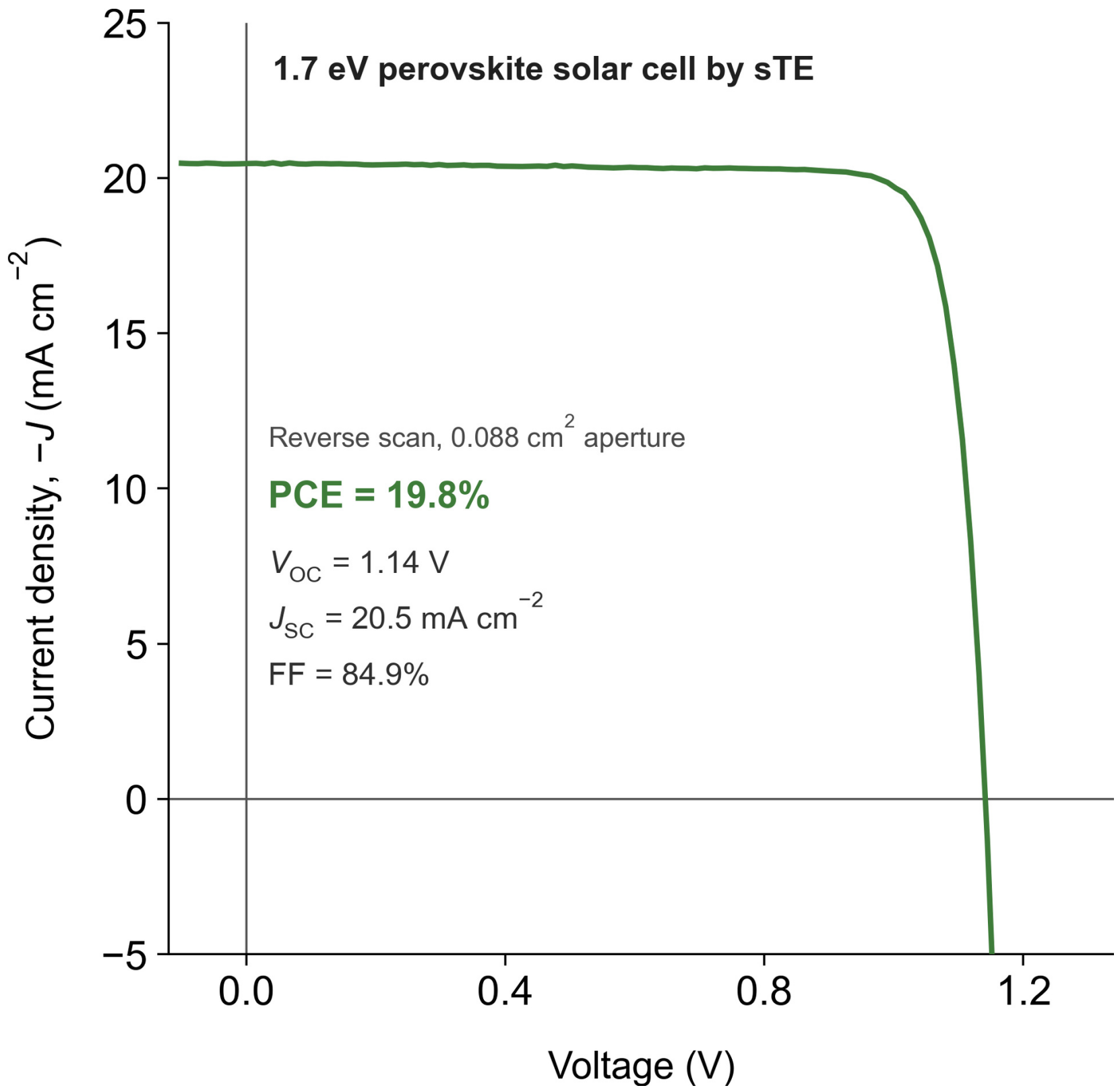

**Supplementary Fig. 1 | Independent measurement of a 1.70-eV sTE perovskite solar cell.**

*J*–*V* characteristic and photovoltaic parameters for a single-junction device measured by the National Photovoltaic Industry Measurement and Testing Center, China, under simulated AM 1.5G illumination through a 0.088-$cm^2$ aperture. The photovoltaic parameters were $J_{SC}$ = 20.5 mA $cm^{-2}$, $V_{OC}$ = 1.14 V, FF = 84.9% and PCE = 19.8%.

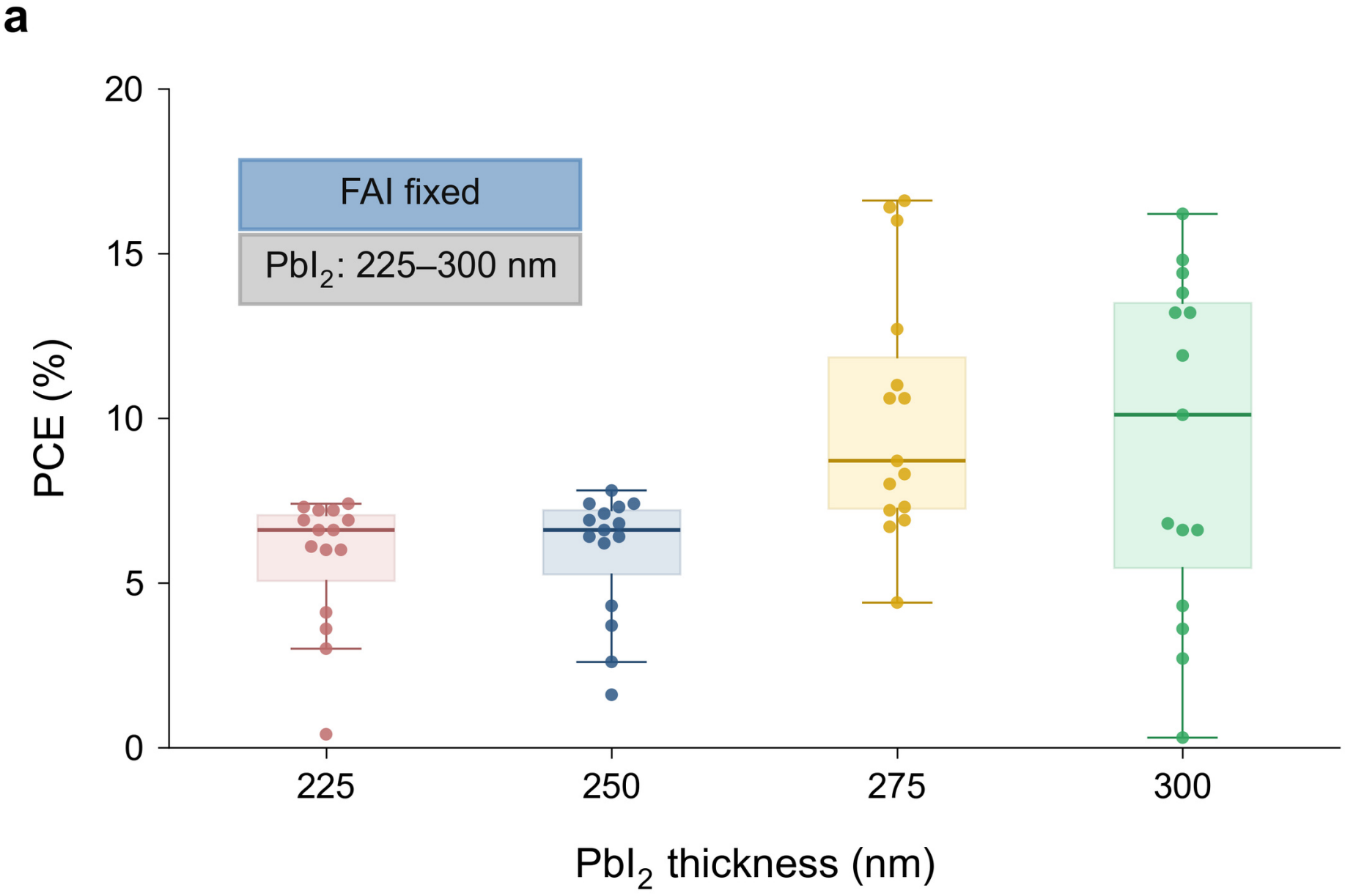

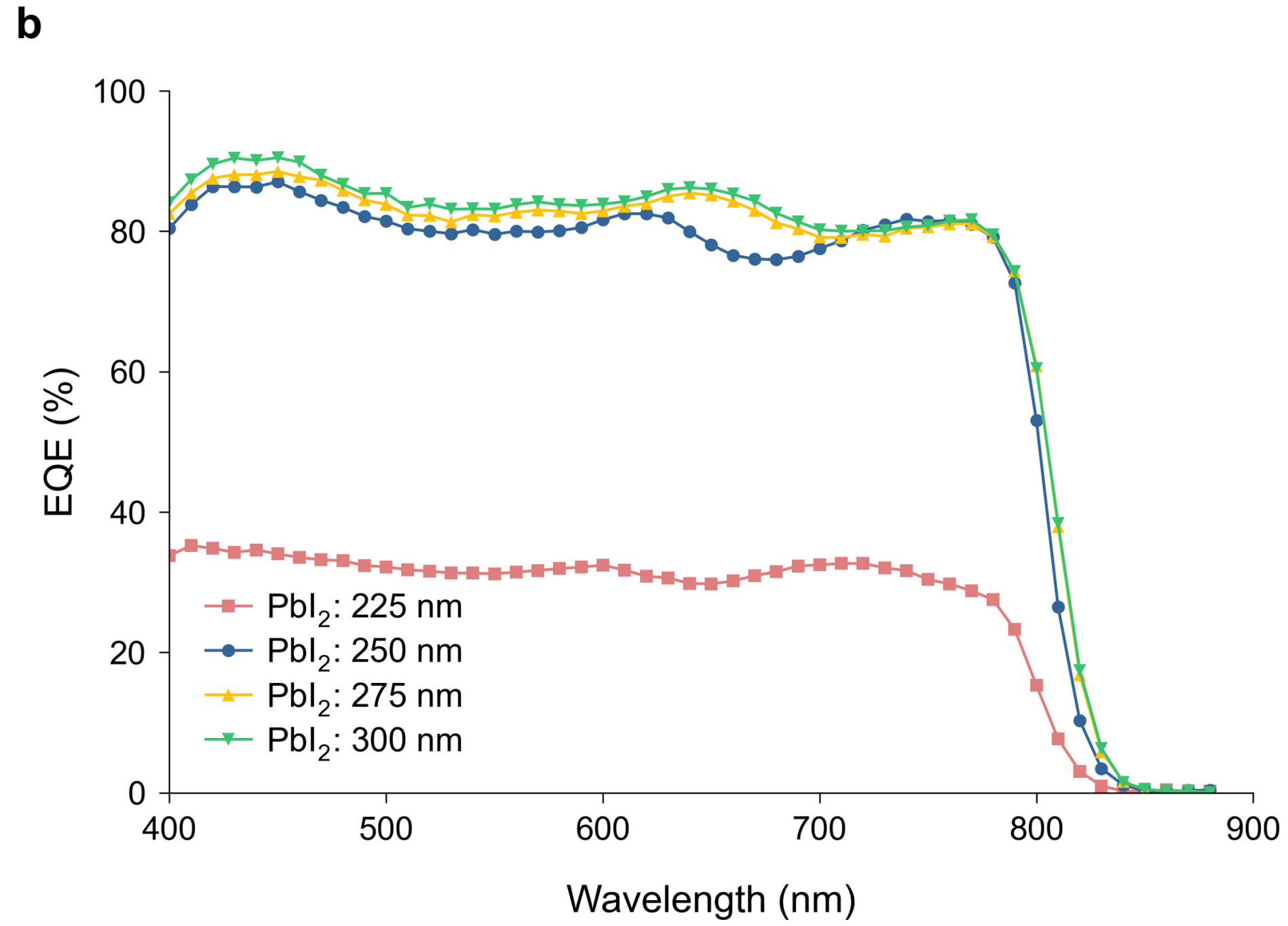

**Supplementary Fig. 2 | PCE distributions and EQE spectra of MBG absorbers prepared with different $PbI_2$ thicknesses.**

**a,** PCE distributions for devices fabricated from $PbI_2$/FAI stacks with $PbI_2$ thicknesses of 225, 250, 275 and 300 nm at a fixed FAI thickness of 350 nm. Individual devices are shown as points; boxes show the interquartile range and median, and whiskers extend to 1.5 times the interquartile range ($n$ = 15 devices per condition). **b,** Corresponding EQE spectra.

**a**

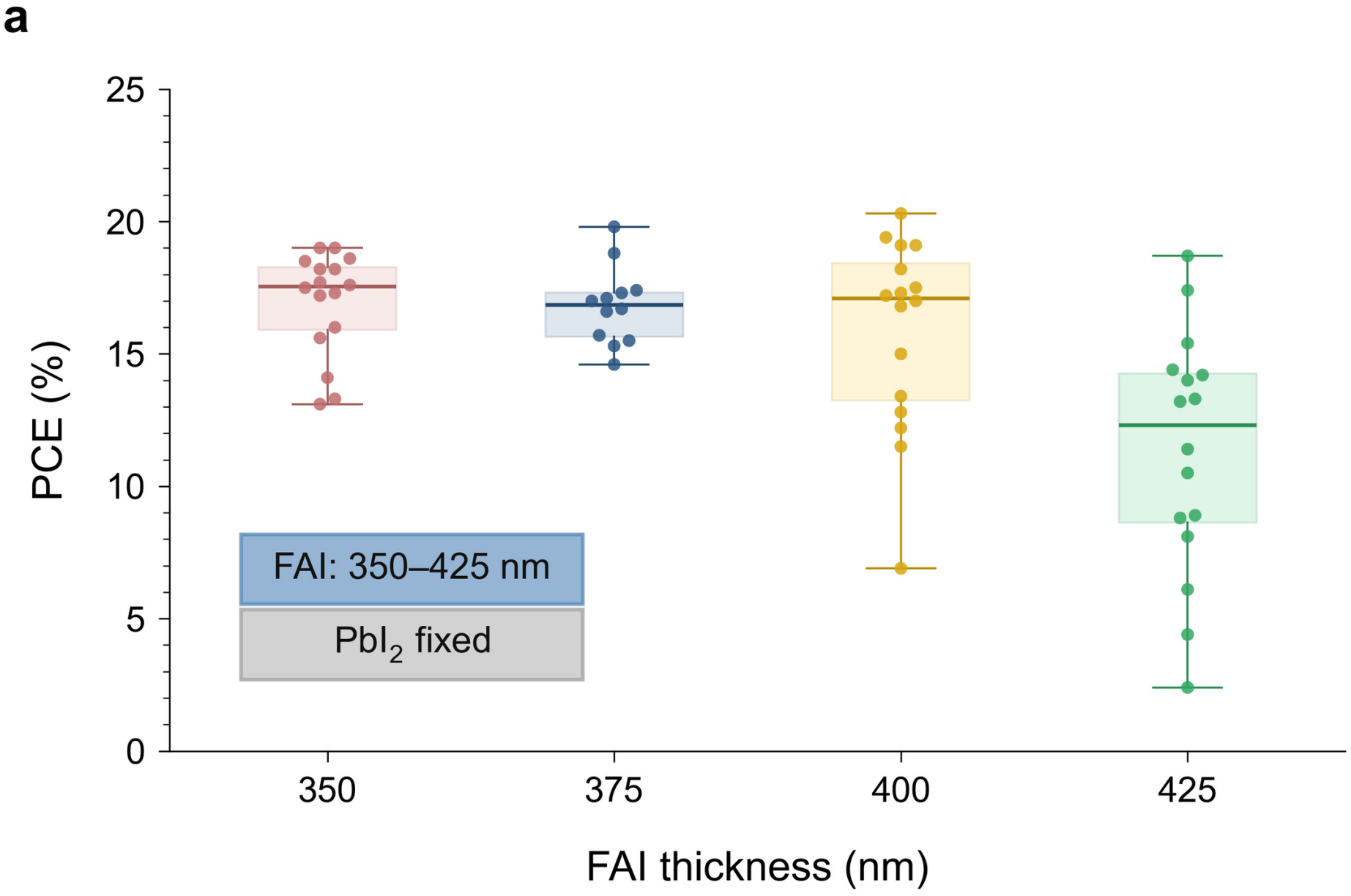

**b**

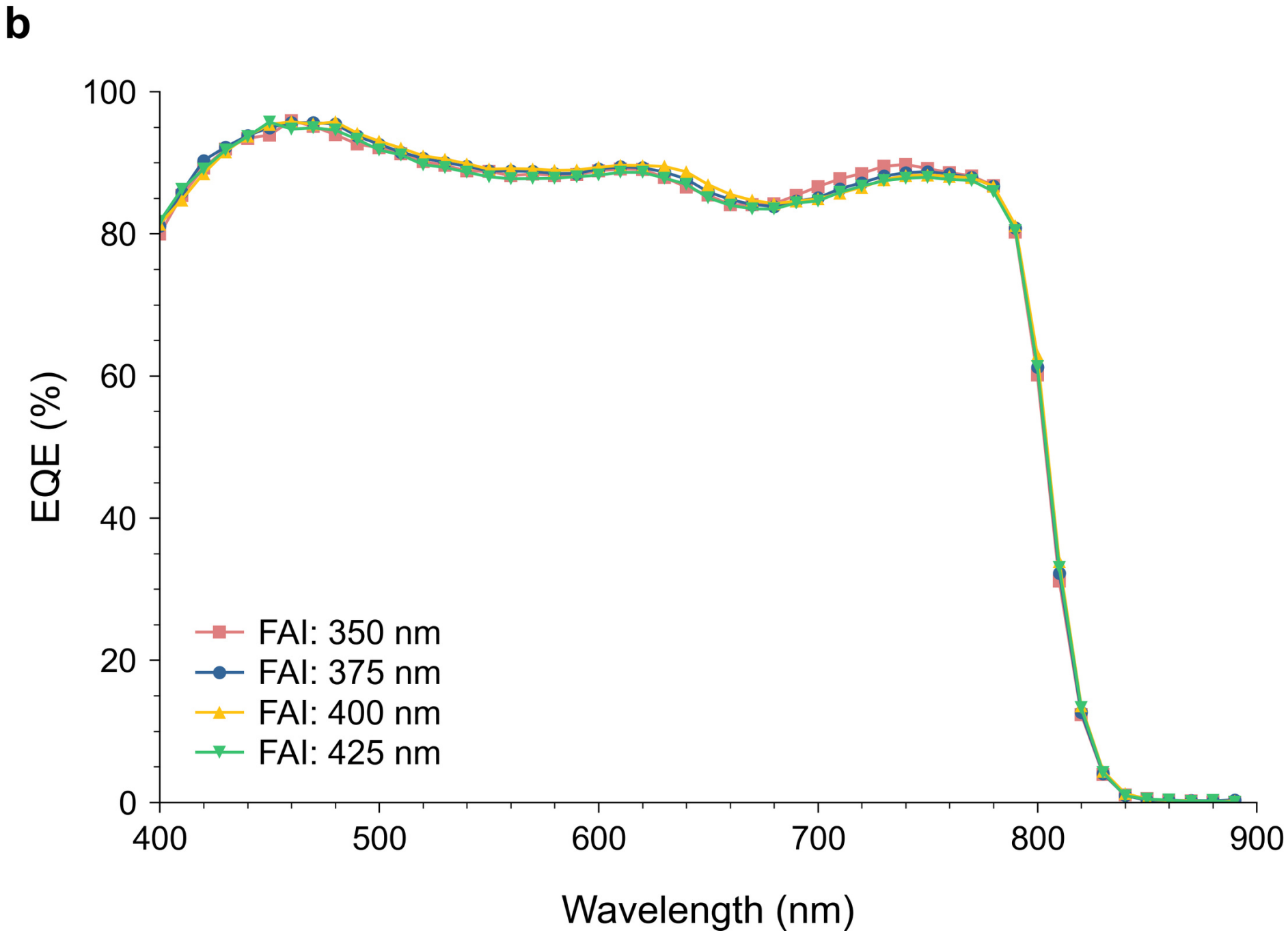

**Supplementary Fig. 3 | PCE distributions and EQE spectra of MBG absorbers prepared with different FAI thicknesses.**

**a,** PCE distributions for devices fabricated from $PbI_2$/FAI stacks with FAI thicknesses of 350, 375, 400 and 425 nm at a fixed $PbI_2$ thickness of 280 nm. Individual devices are shown as points; boxes show the interquartile range and median, and whiskers extend to 1.5 times the interquartile range ($n$ = 16, 12, 16 and 16 devices for FAI thicknesses of 350, 375, 400 and 425 nm, respectively). **b,** Corresponding EQE spectra.

a

b

| $PbBr_2$ (nm) | $PbI_2$ (nm) | Total (nm) | $PbBr_2/(PbI_2 + PbBr_2)$ (%) | $E_g$ from EQE (eV) |
|---|---|---|---|---|
| 150 | 130 | 280 | 53.57 | 1.785 |
| 160 | 140 | 300 | 53.33 | 1.795 |
| 170 | 150 | 320 | 53.13 | 1.824 |
| 180 | 160 | 340 | 52.94 | 1.838 |

**Supplementary Fig. 4 | Absolute $PbBr_2$ thickness dominates over nominal inorganic-layer fraction.**

**a,** EQE spectra of WBG devices fabricated with $PbBr_2/PbI_2$ thickness pairs of 150/130, 160/140, 170/150 and 180/160 nm at fixed FAI and CsI thicknesses. **b,** Nominal precursor thicknesses, $PbBr_2$ thickness fraction and EQE-derived bandgap for the stacks in **a**. The bandgap increases from 1.785 to 1.838 eV as the absolute $PbBr_2$ thickness increases, although the nominal $PbBr_2$ fraction decreases from 53.6% to 52.9%.

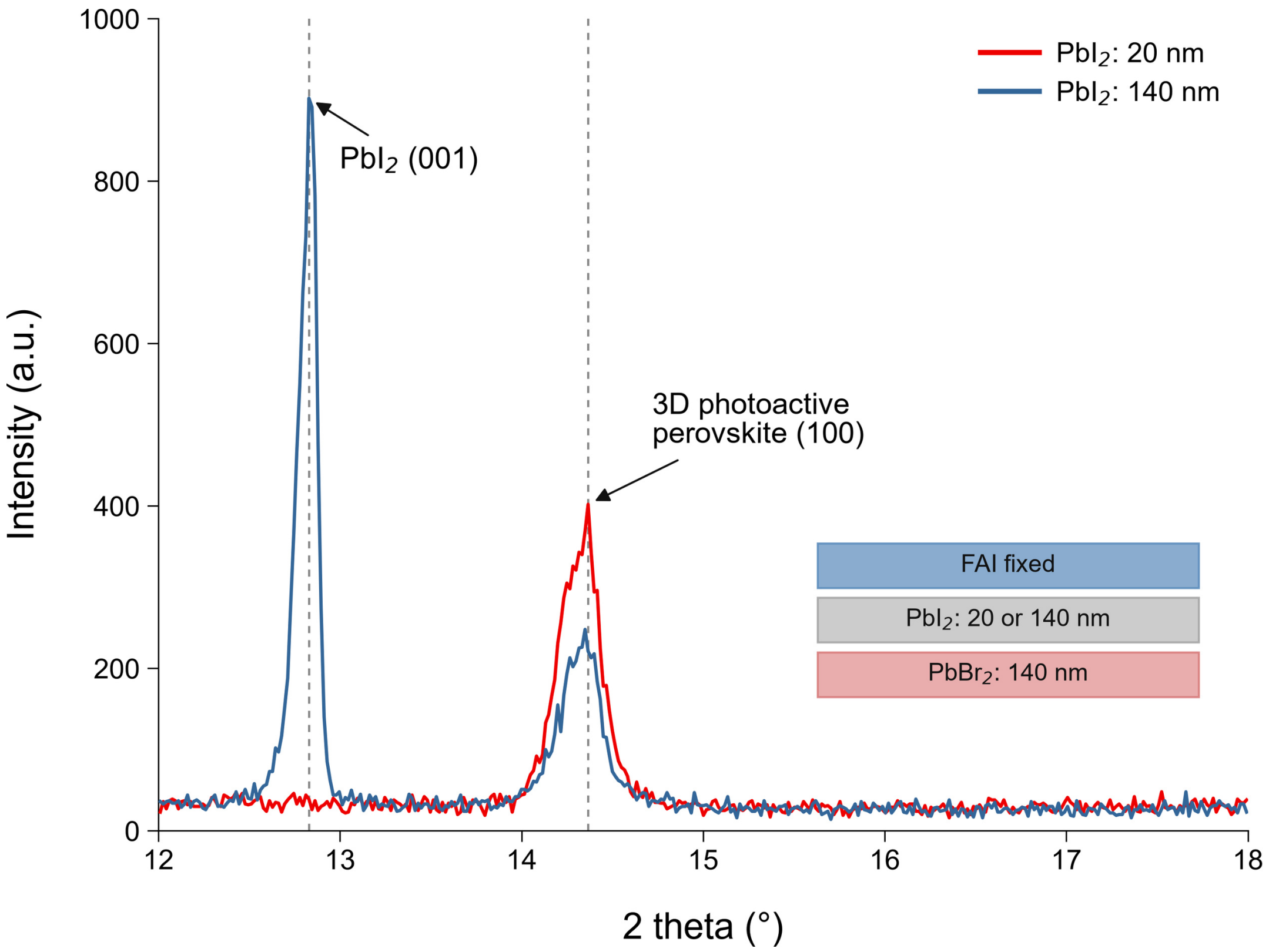

**Supplementary Fig. 5 | Excess $PbI_2$ remains as a residual inorganic phase in the WBG absorber.**

Ex situ XRD patterns of annealed $PbBr_2/PbI_2$/FAI/CsI stacks prepared with 20- or 140-nm $PbI_2$ at a fixed $PbBr_2$ thickness of 140 nm and otherwise identical conditions. The $PbI_2$ (001) reflection near 12.8° is prominent in the 140-nm $PbI_2$ group, whereas the perovskite reflection remains near 14.4°. Together with the nearly invariant EQE onset, this result indicates that added $PbI_2$ is not proportionally incorporated into the photoactive phase.

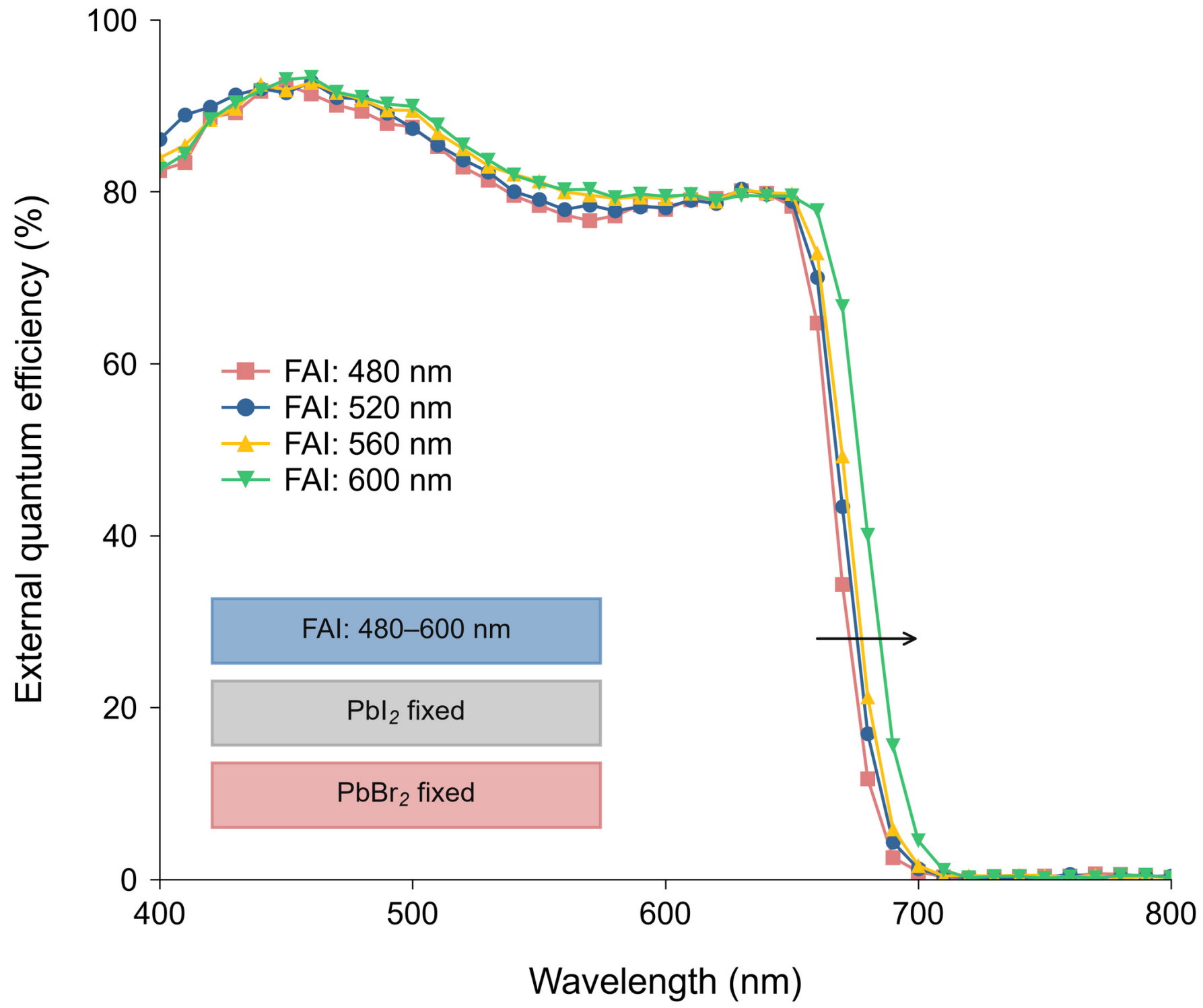

**Supplementary Fig. 6 | Weak bandgap response to FAI thickness in the WBG absorber.**

EQE spectra of representative WBG devices fabricated with FAI thicknesses of 480, 520, 560 and 600 nm, at fixed $PbBr_2$, $PbI_2$ and CsI. Increasing FAI shifts the EQE-derived bandgap only from 1.854 to 1.822 eV while leaving the visible-range response largely unchanged.

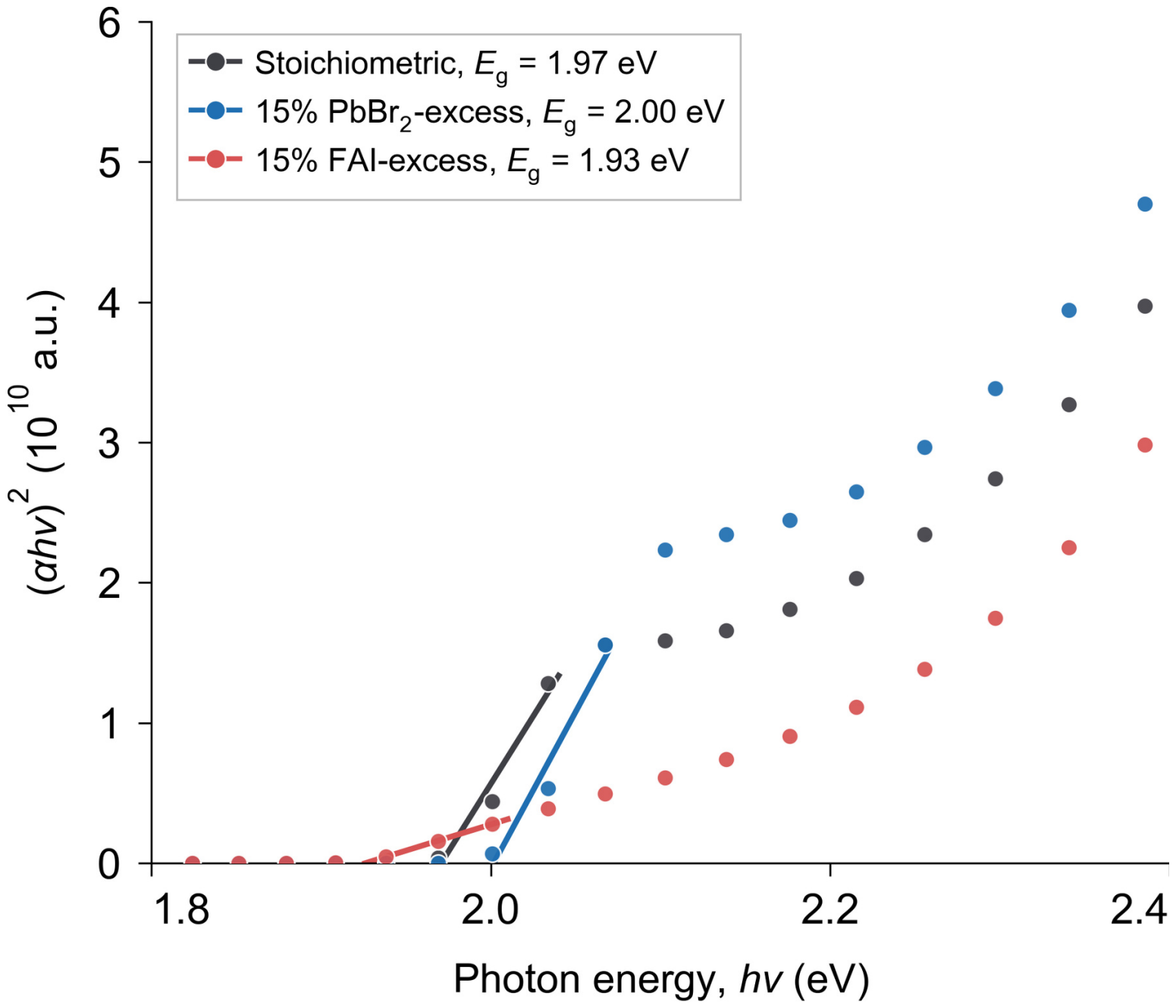

**Supplementary Fig. 7 | Optical bandgaps of spin-coated $PbBr_2$/FAI reference films.**

Direct-transition Tauc plots of spin-coated films prepared with stoichiometric, 15% $PbBr_2$-excess and 15% FAI-excess compositions, each containing 5 mol% CsI. The extracted bandgaps are 1.97, 2.00 and 1.93 eV, respectively.

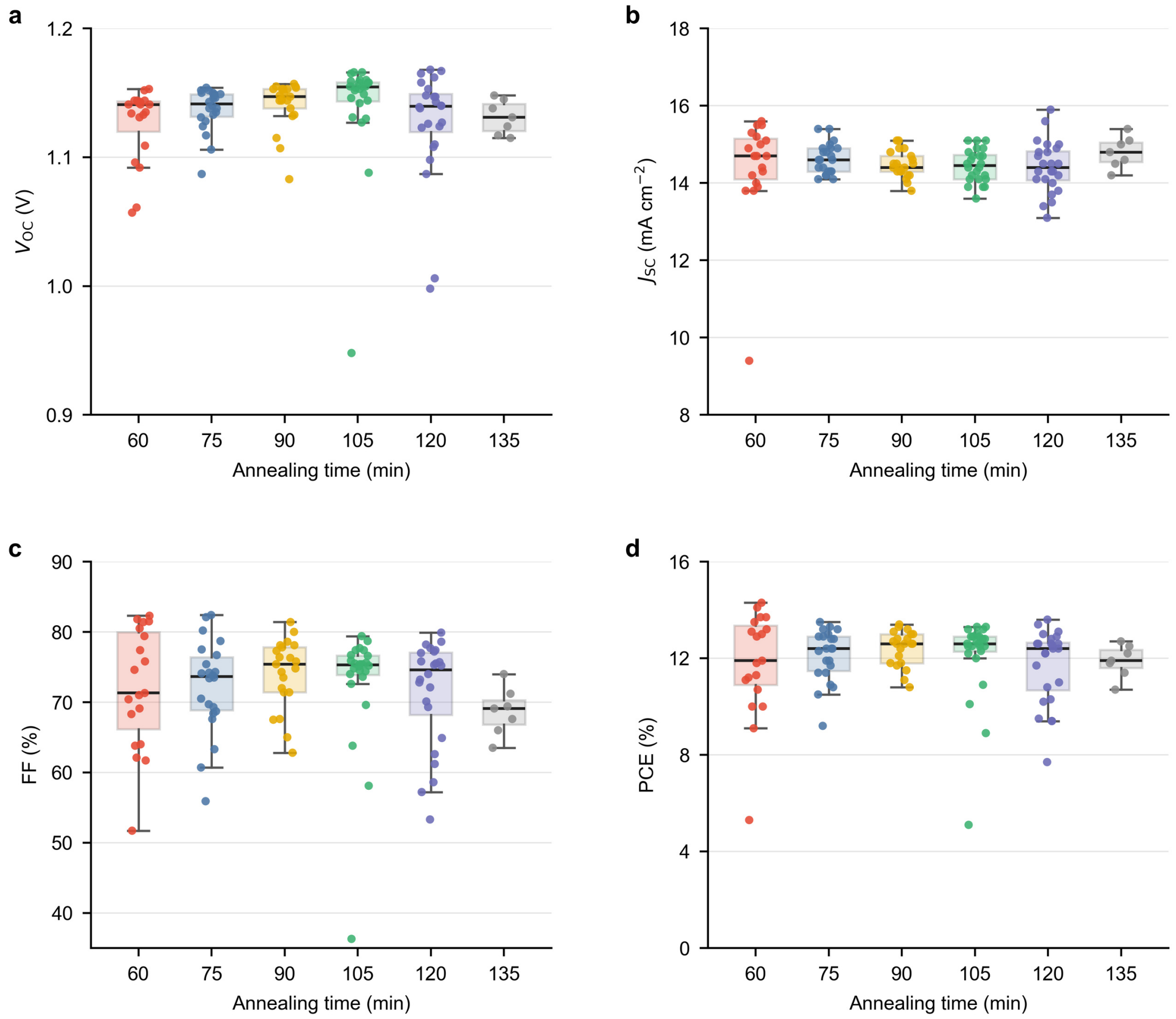

**Supplementary Fig. 8 | A thin $PbI_2$ layer broadens the annealing window of the WBG stack.**

Distributions of **a,** $V_{OC}$, **b,** $J_{SC}$, **c,** FF and **d,** PCE for WBG devices with $PbBr_2/PbI_2$/FAI/CsI stack, containing a 20-nm $PbI_2$ layer and annealed in ambient air at 170 °C and 30–40% relative humidity for 60–135 min. Individual devices are shown as points; boxes show the interquartile range and median, and whiskers extend to 1.5 times the interquartile range. $n$ = 19, 22, 21, 24, 24 and 7 devices for 60, 75, 90, 105, 120 and 135 min, respectively.

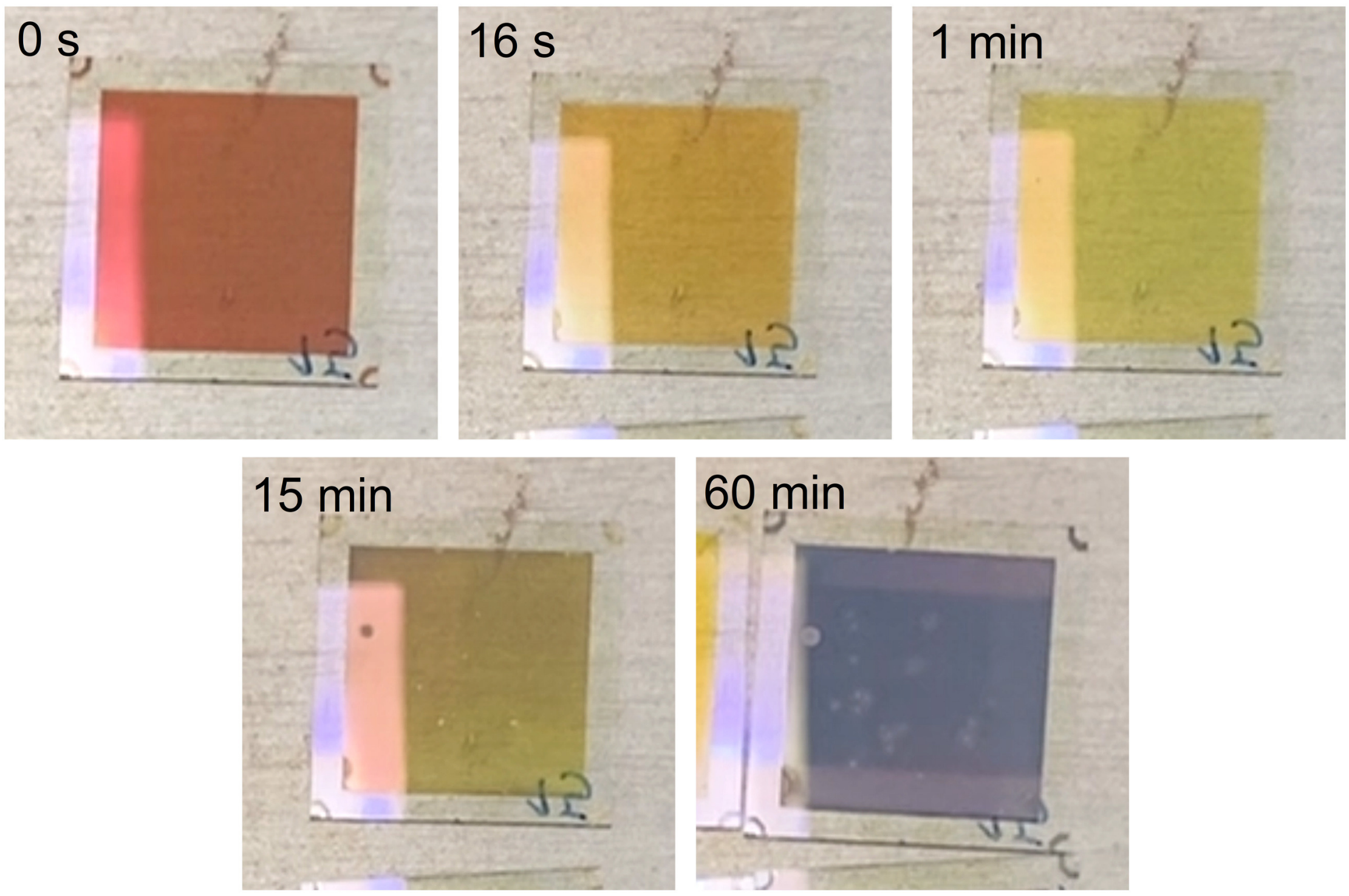

**Supplementary Fig. 9 | Slow darkening of WBG sTE stacks during annealing.**

Time-series photographs of $PbBr_2/PbI_2$/FAI/CsI stacks annealed in ambient air at 170 °C and 30–40% relative humidity. Time labels are referenced to the first frame. The reddish as-deposited stack becomes yellow within approximately 1 min and develops a dark perovskite-like appearance only after prolonged annealing. The corresponding recording is Supplementary Video 1.

**Supplementary Video 1 | Slow darkening of WBG sTE stacks during annealing.** Time-lapse video of a $PbBr_2/PbI_2$/FAI/CsI stack annealed in ambient air at 170 °C; the recording corresponds to Supplementary Fig. 9.

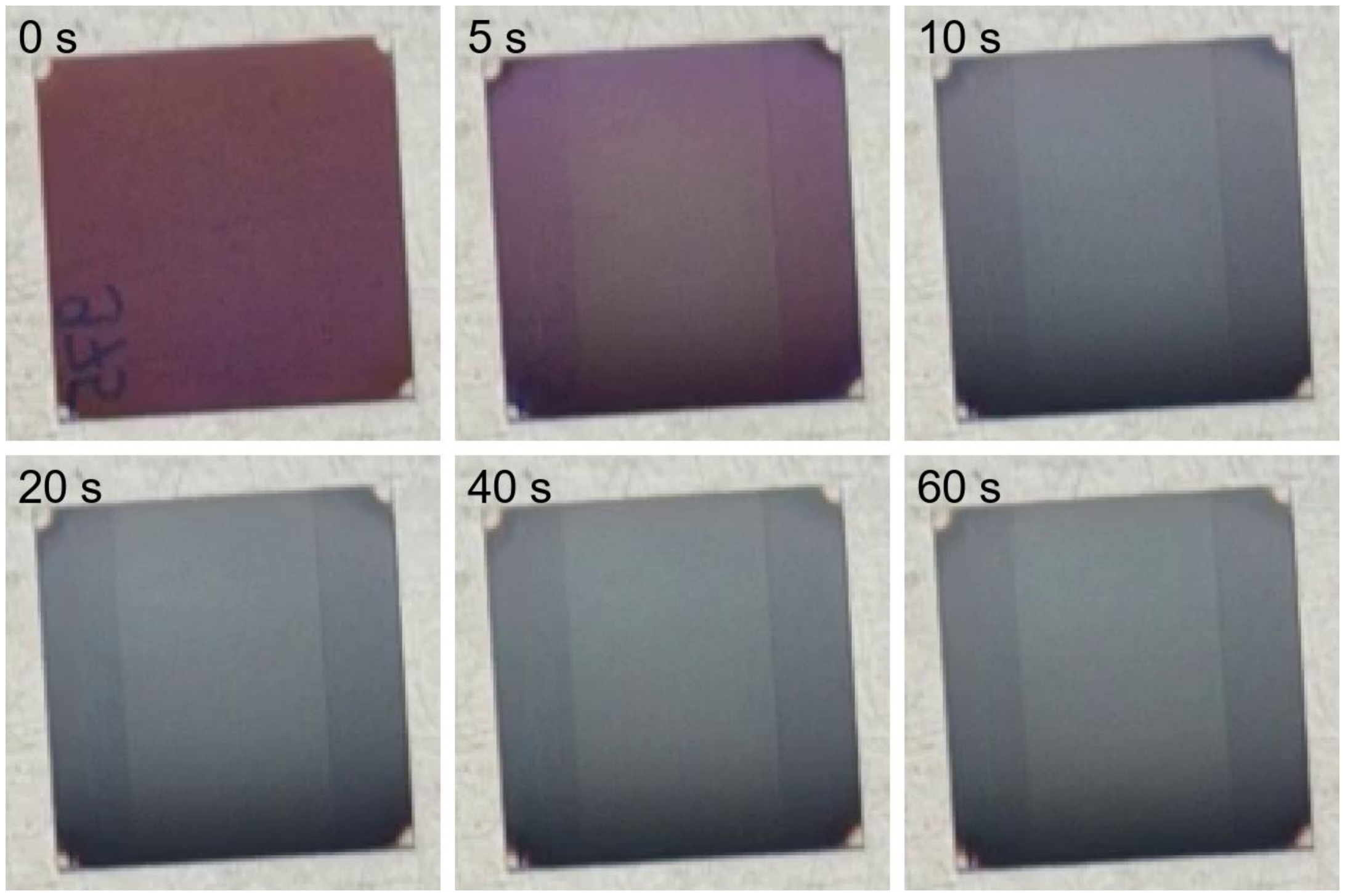

**Supplementary Fig. 10 | Rapid darkening of the MBG $PbI_2$/FAI reference.**

Time-series photographs of $PbI_2$/FAI stacks annealed in ambient air at 170 °C and 30–40% relative humidity. Time labels are referenced to the first frame. The stack darkens rapidly to a perovskite-like appearance within approximately 10 s; later frames through 60 s show little further visible change. The corresponding recording is Supplementary Video 2.

**Supplementary Video 2 | Rapid darkening of the MBG stack during annealing.** Time-lapse video of a $PbI_2$/FAI stack annealed in ambient air at 170 °C from 0 to 60 s; the recording corresponds to Supplementary Fig. 10.

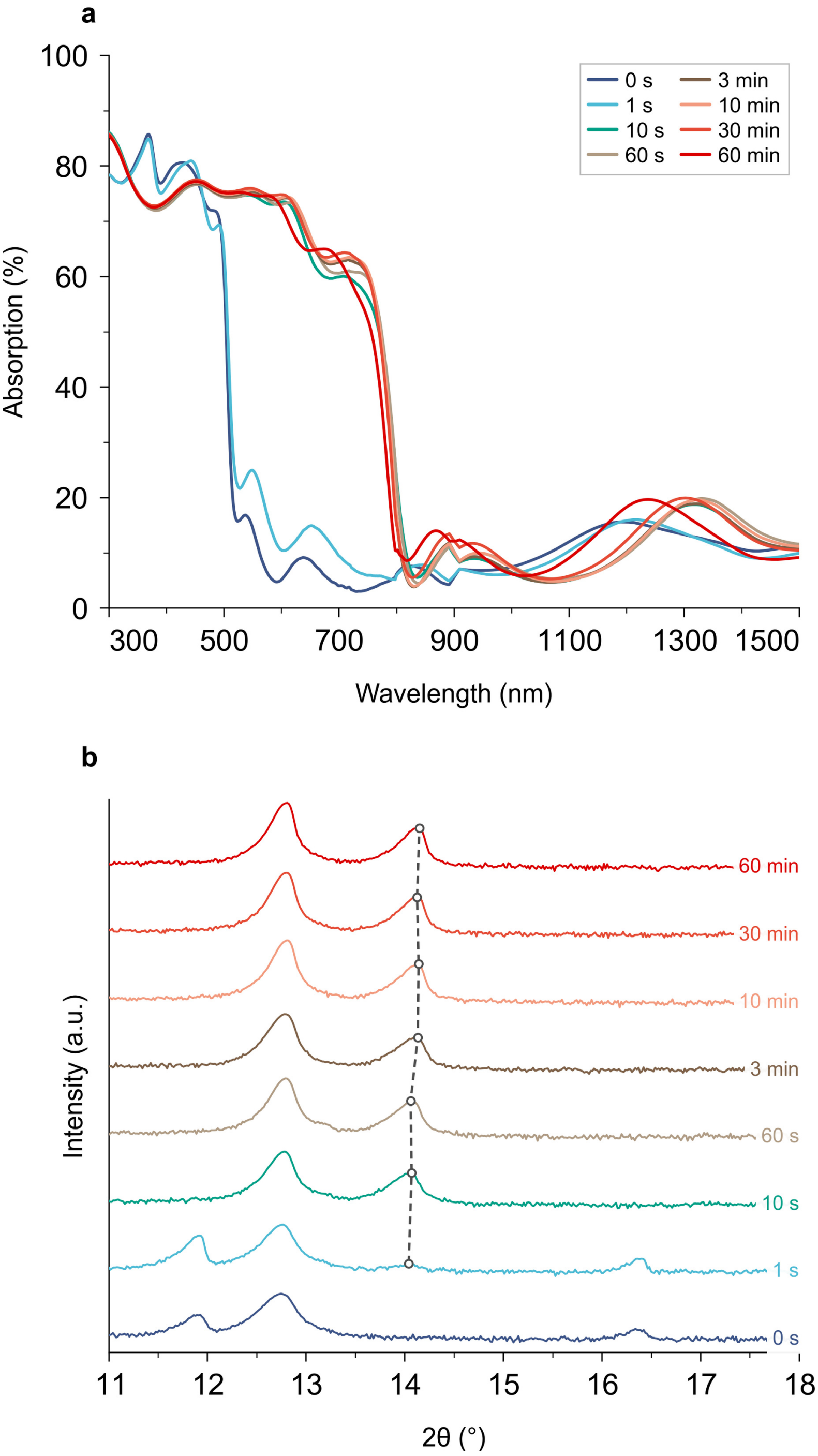

**Supplementary Fig. 11 | Rapid optical and structural development of the MBG stack.**

**a**, UV–vis absorption spectra of the MBG $PbI_2$/FAI stacks annealed for the indicated durations. **b**, X-ray diffraction patterns of the MBG $PbI_2$/FAI stacks annealed for the indicated durations. Open circles mark the maximum of the perovskite reflection (~14°) from 1 s onwards, and the dashed line is a guide to the eye.

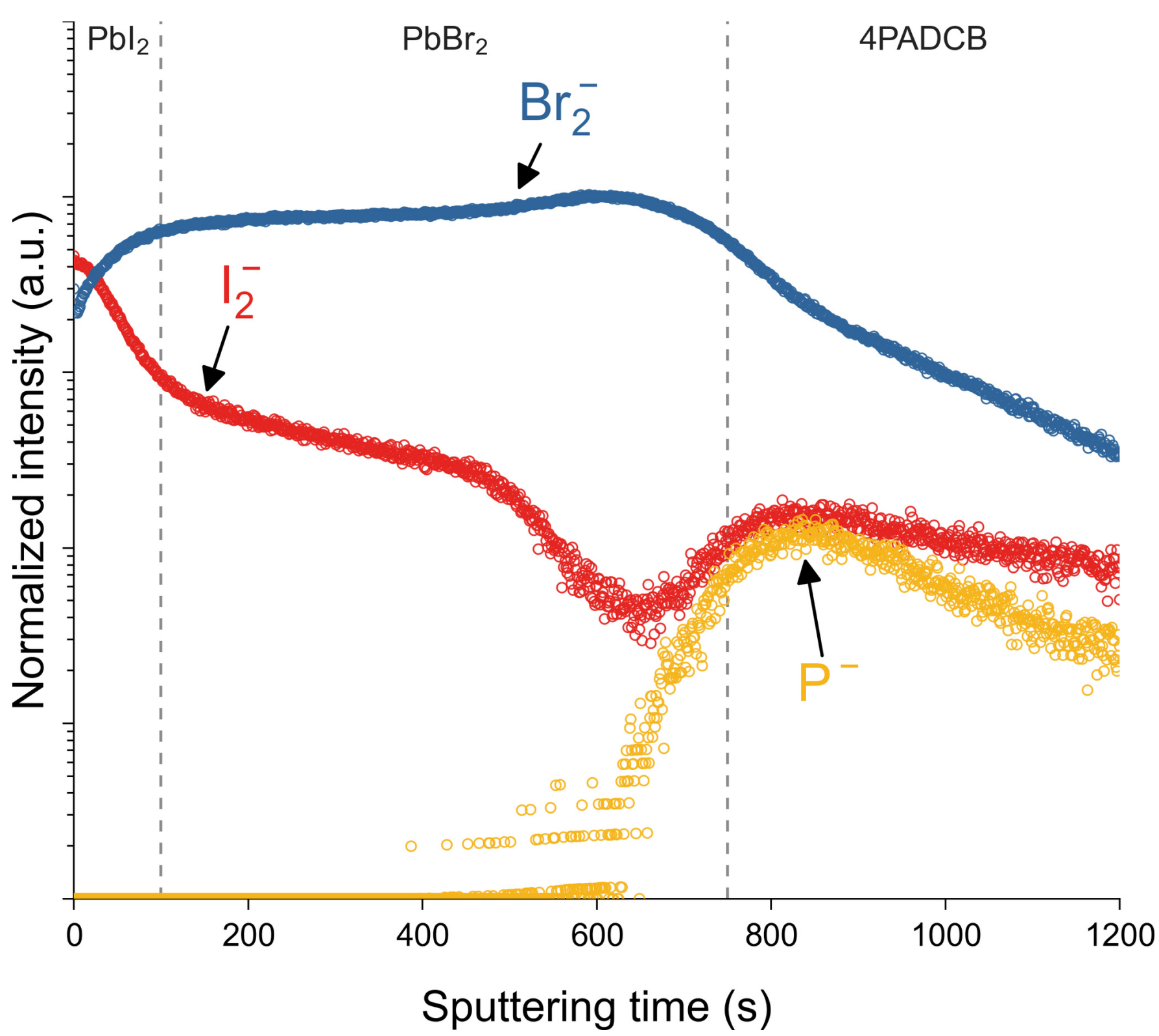

**Supplementary Fig. 12 | The WBG inorganic template is chemically stratified before FAI deposition.**

ToF-SIMS depth profiles of the as-deposited ITO/4PADCB/$PbBr_2$/$PbI_2$ template. $I_2^-$, $Br_2^-$ and $P^-$ track $PbI_2$-, $PbBr_2$- and SAM-containing regions, respectively.

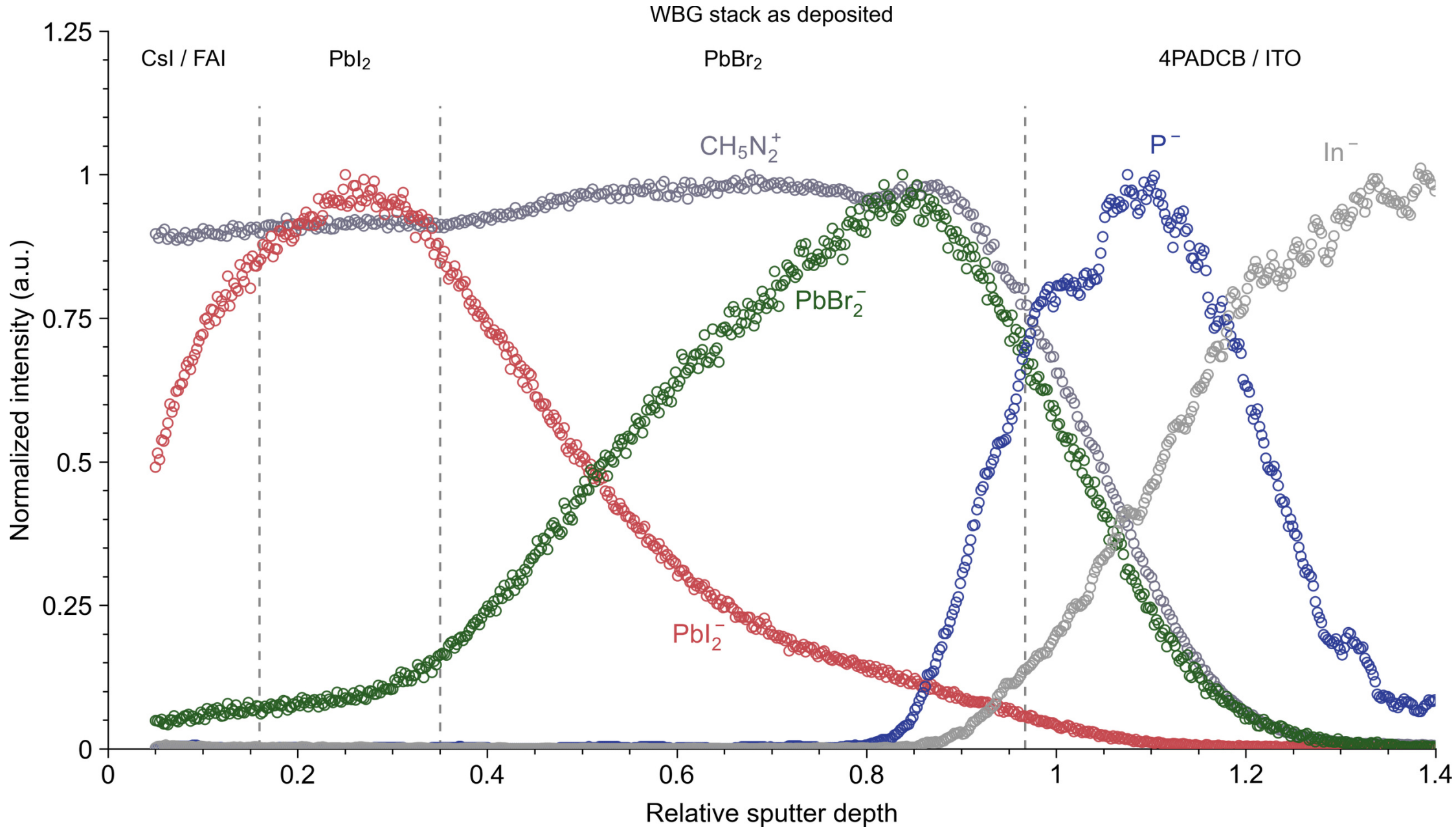

**Supplementary Fig. 13 | FA-containing species penetrate the WBG inorganic region before annealing.**

Positive-ion ($CH_5N_2^+$) and negative-ion ($PbI_2^-$, $PbBr_2^-$, $P^-$ and $In^-$) ToF-SIMS profiles of the as-deposited ITO/4PADCB/$PbBr_2$/$PbI_2$/FAI/CsI stack. Profiles of opposite polarity are aligned on a common normalized depth coordinate using the $In^-$ onset. Dashed lines indicate chemically assigned regions. The $CH_5N_2^+$ signal extends across the inorganic layer while $PbI_2^-$ and $PbBr_2^-$ remain depth dependent.

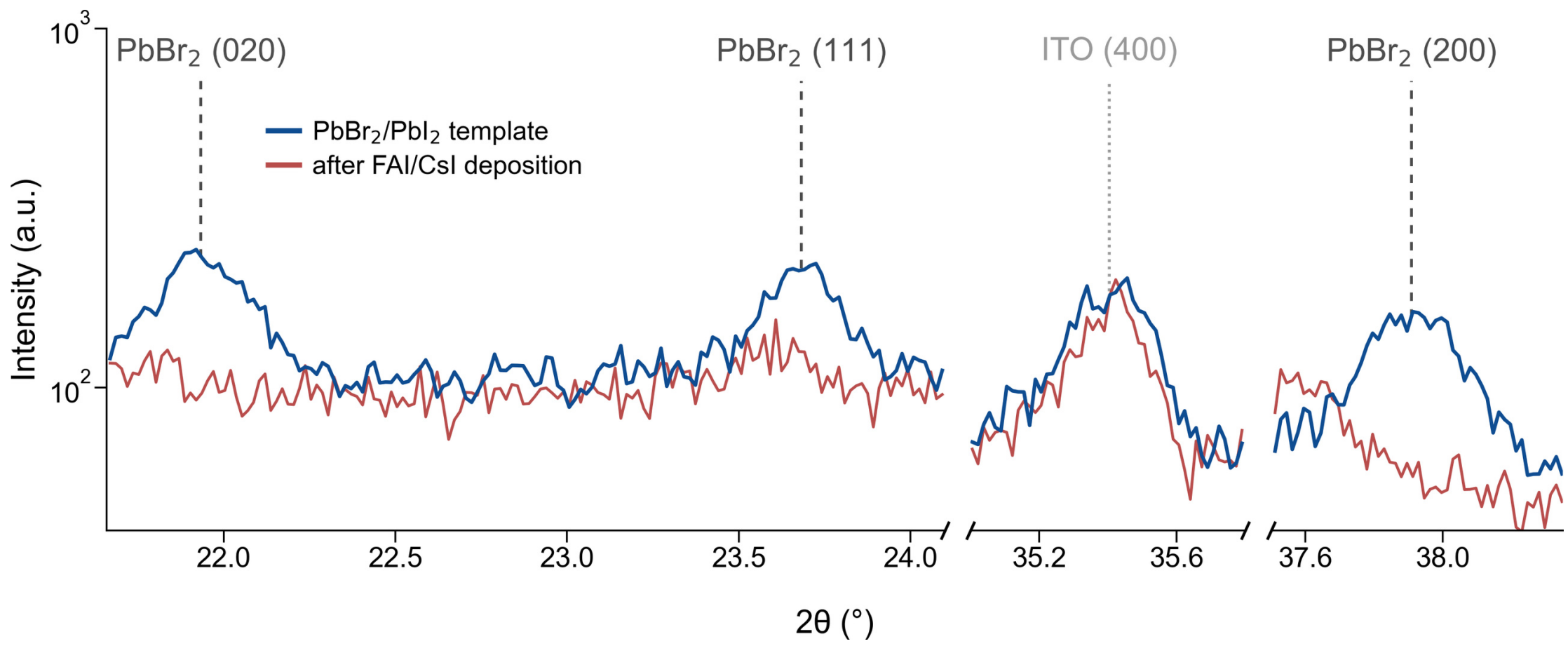

**Supplementary Fig. 14 | Crystalline $PbBr_2$ is not detected in the as-deposited $PbBr_2/PbI_2$/FAI/CsI stack.**

XRD patterns of the $PbBr_2/PbI_2$ inorganic template and $PbBr_2/PbI_2$/FAI/CsI stack. Dashed guides mark the $PbBr_2$ (020), (111) and (200) reflections using $In_2O_3$ (400) as an internal reference. $PbBr_2$ reflections are resolved only in the inorganic template; no crystalline $PbBr_2$ remains detectable after FAI/CsI deposition.[31]

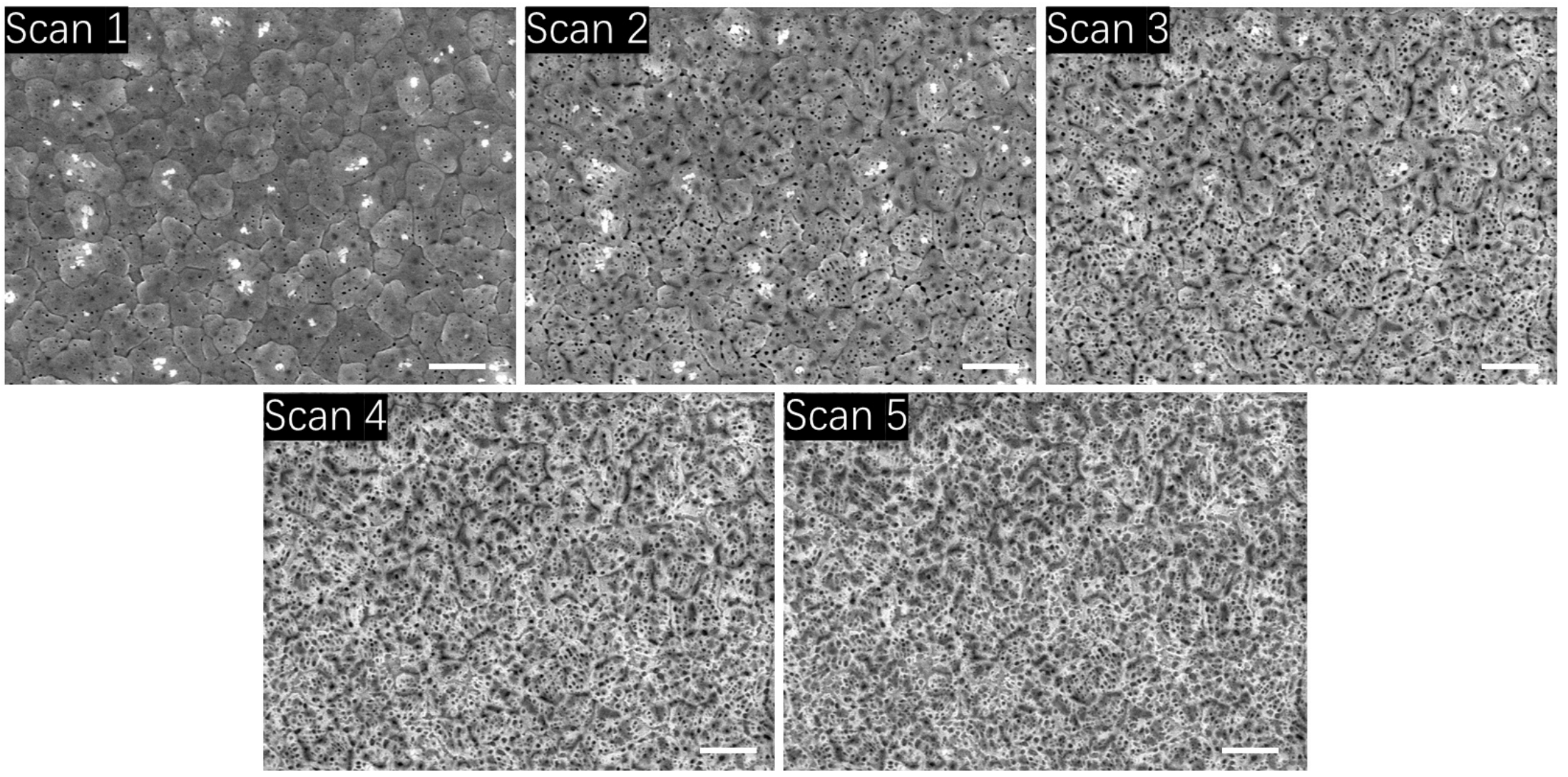

**Supplementary Fig. 15 | Electron-beam sensitivity of the as-deposited WBG stack surface.**

Consecutive top-view SEM scans of the same region of an as-deposited ITO/4PADCB/$PbBr_2$/$PbI_2$/FAI/CsI stack at 5 kV. Progressive contrast and morphology changes reveal pronounced beam sensitivity, consistent with an organic-halide-rich precursor surface. Scale bars, 1 μm.

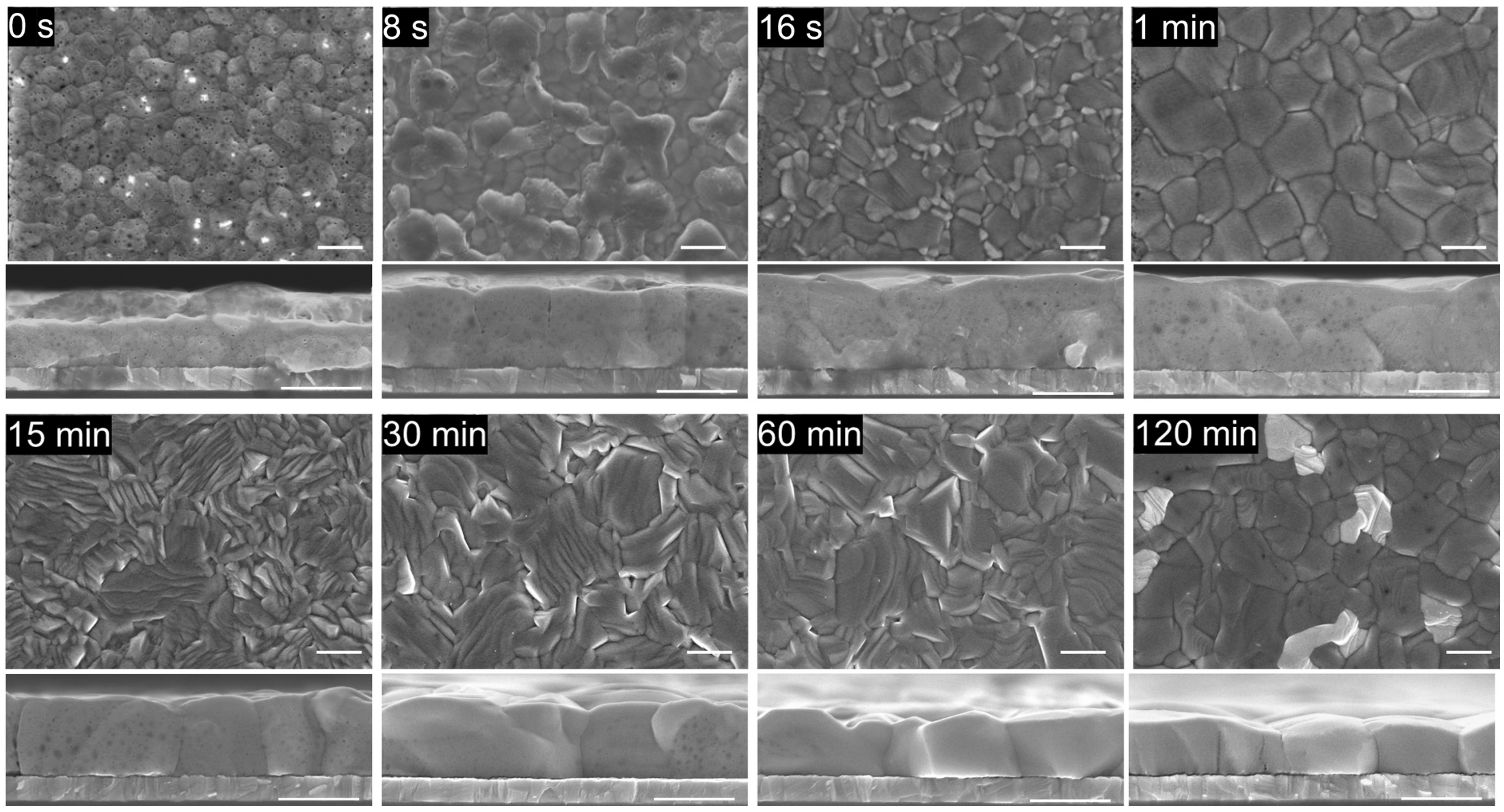

**Supplementary Fig. 16 | Morphological evolution of WBG stacks during annealing.**

Top-view and cross-sectional SEM images of the ITO/4PADCB/$PbBr_2$/$PbI_2$/FAI/CsI stacks after the annealing times indicated above the panels; each image was acquired from a separate sample. The surface reorganizes rapidly and the film initially expands, followed by grain reconstruction, densification and thickness reduction during prolonged annealing. Scale bars, 1 μm (top views) and 500 nm (cross-sections).

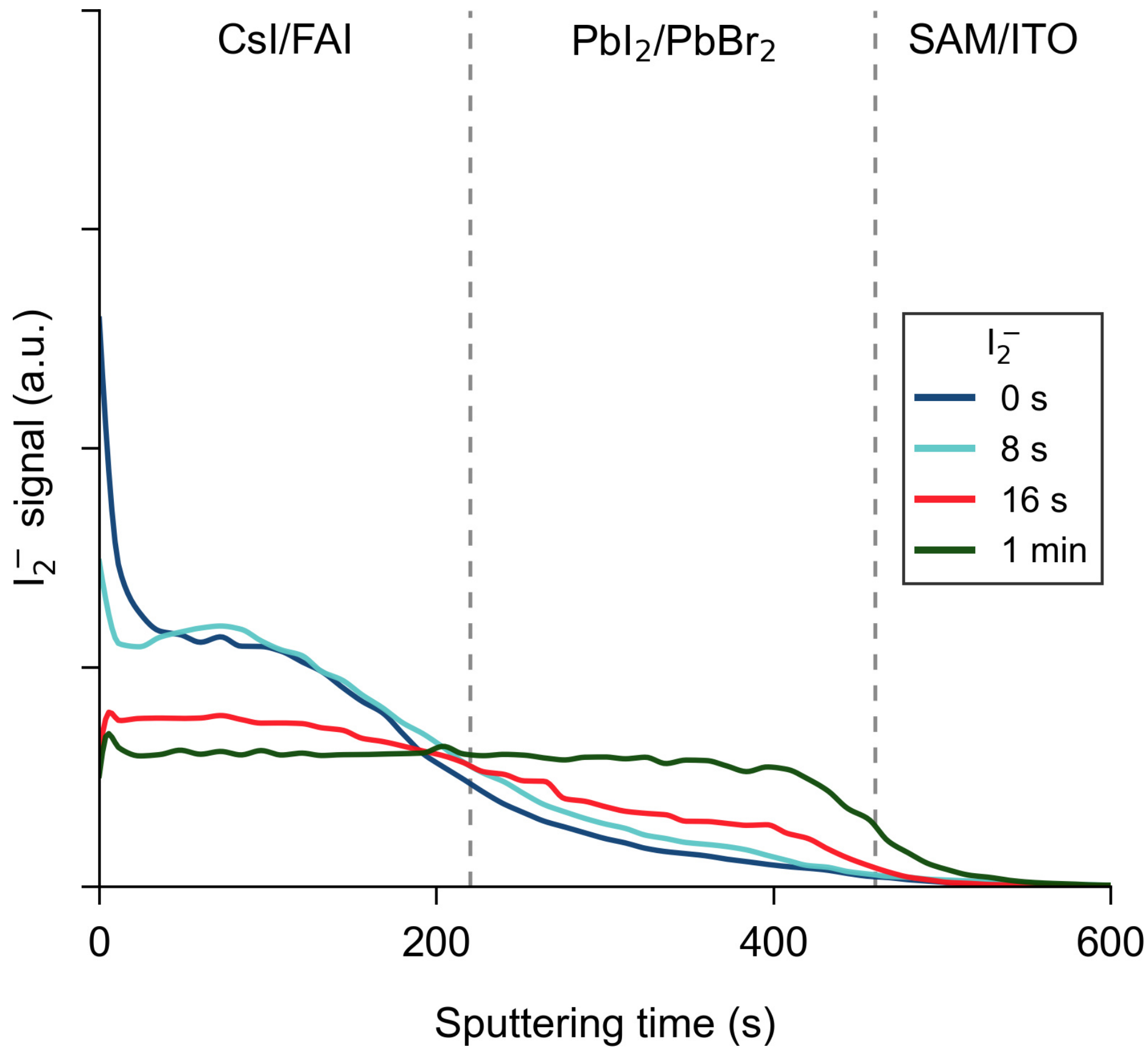

**Supplementary Fig. 17 | Iodide redistribution during the first minute of annealing for the WBG stack.**

$I_2^-$ ToF-SIMS depth profiles of the ITO/4PADCB/$PbBr_2$/$PbI_2$/FAI/CsI stacks after the indicated annealing times from 0 to 60 s. The initially depth-dependent iodide distribution becomes substantially flatter within 60 s, similar to the $Br_2^-$ evolution in **Fig. 3c**.

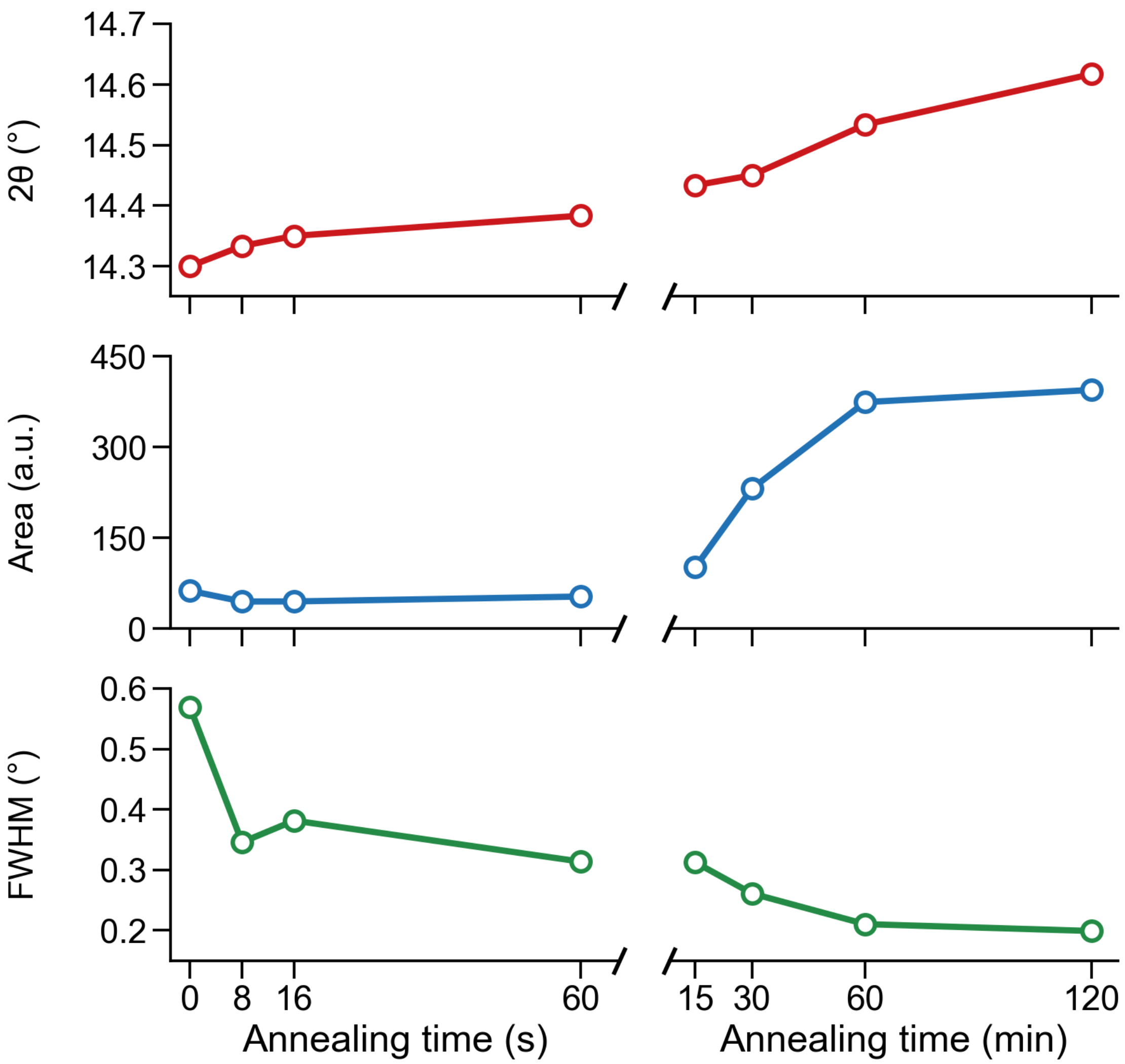

**Supplementary Fig. 18 | XRD peak parameters during WBG conversion.**

Top to bottom, peak position, integrated area and full-width at half-maximum (FWHM) of the perovskite-related diffraction feature near 14.3°, extracted from ex situ XRD results. The time axis is broken between 60 s and 15 min.

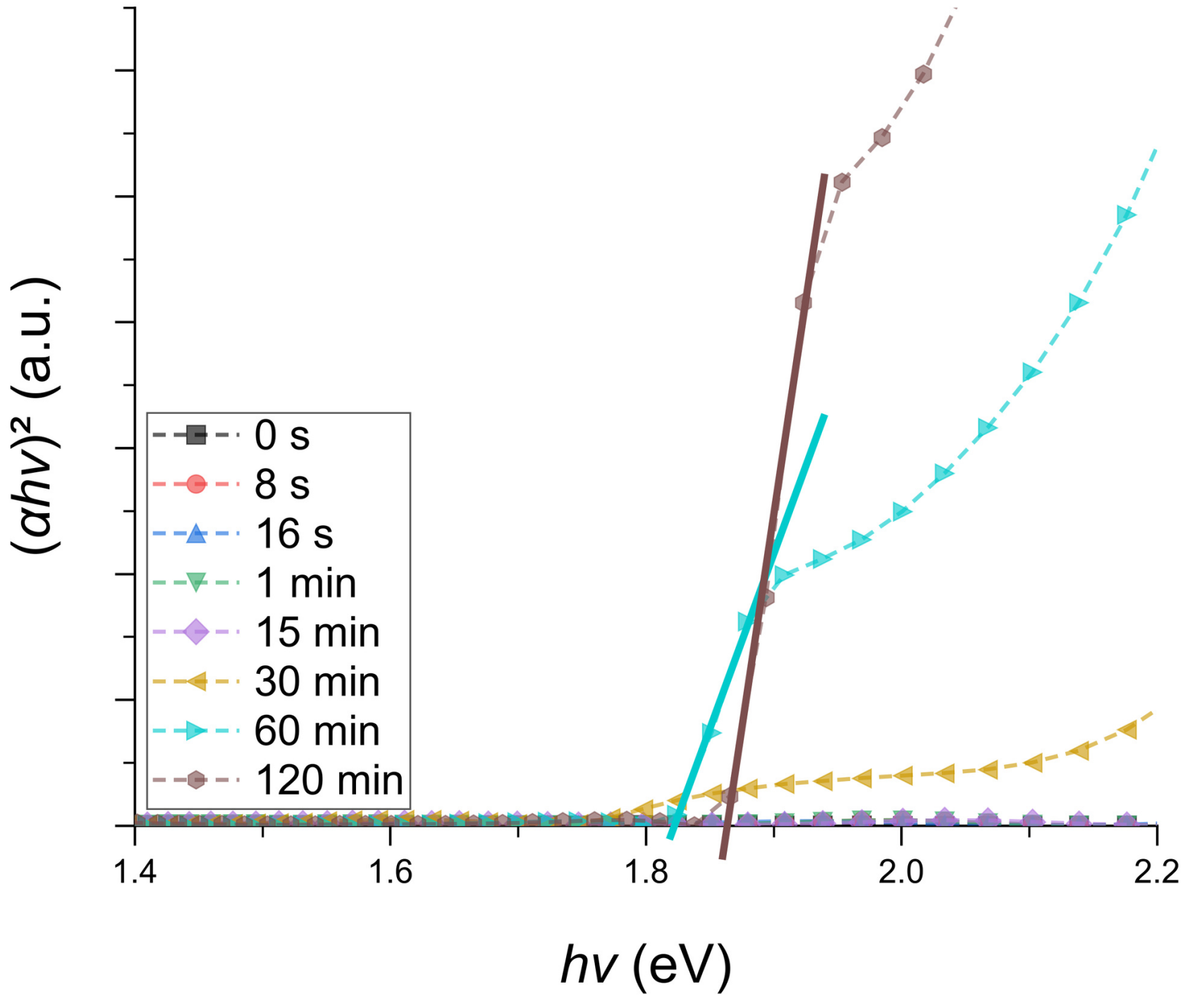

**Supplementary Fig. 19 | A detectable perovskite-like absorption edge emerges only after prolonged annealing.**

Direct-transition Tauc plots derived from transmittance and reflectance spectra of WBG stacks after the indicated annealing times. For the as-deposited and briefly annealed stacks, no optically detectable perovskite-like absorption edge is resolved within the measured spectral range. A clear absorption edge emerges after 30 min and becomes progressively sharper between 60 and 120 min.

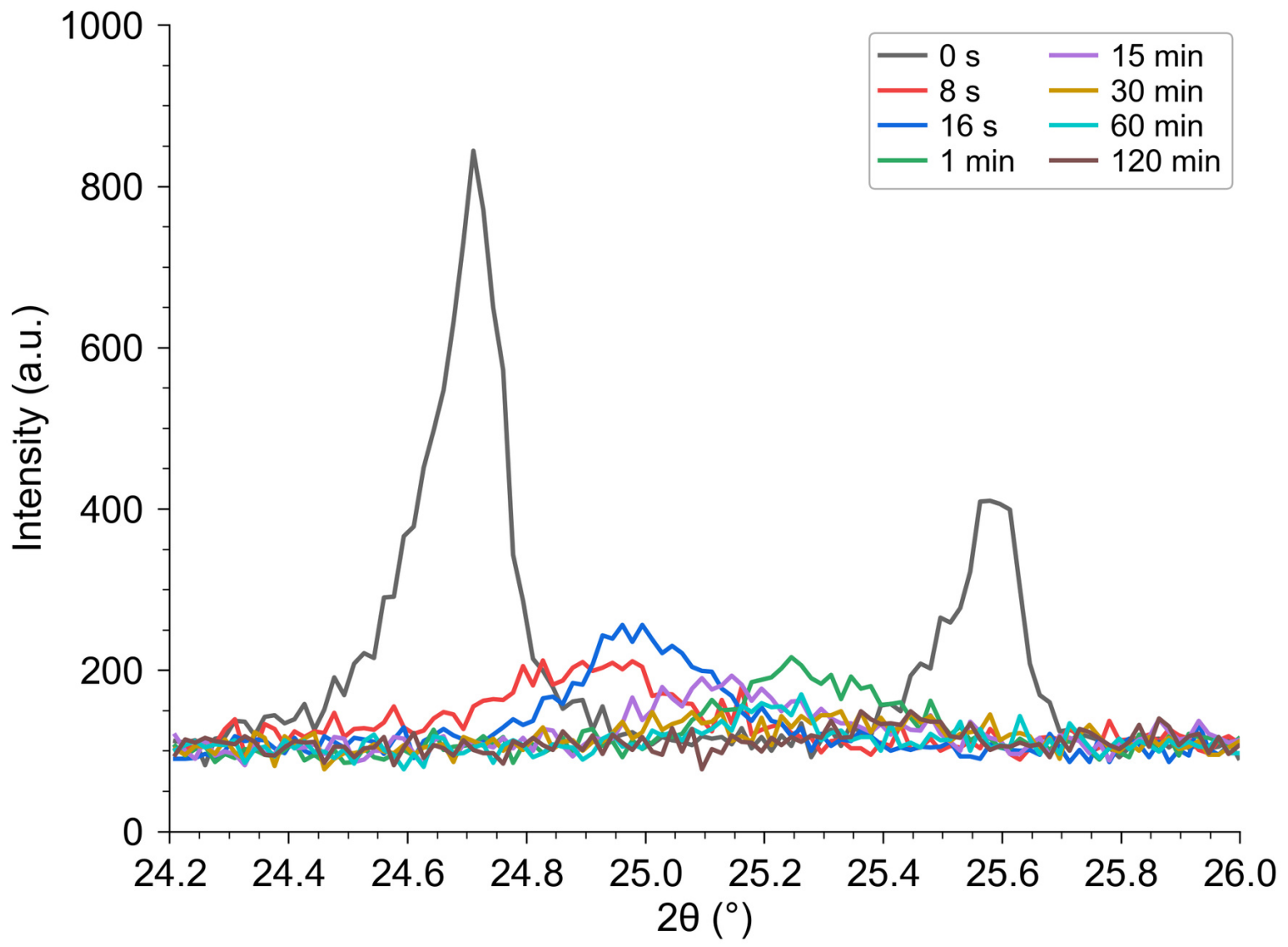

**Supplementary Fig. 20 | Rapid reconstruction of FAI-related diffraction features.**

Raw XRD patterns of WBG stacks in the 24.2–26.0° 2θ range. The reflections at 24.76° and 25.63°, assigned to the ($\bar{1}21$) and (002) planes of monoclinic FAI, collapse within the first 8 s and evolve into a broad organic-halide-related feature. Weak residual intensity at longer annealing times indicates that intermediate character persists during conversion to the 3D phase.[32,33]

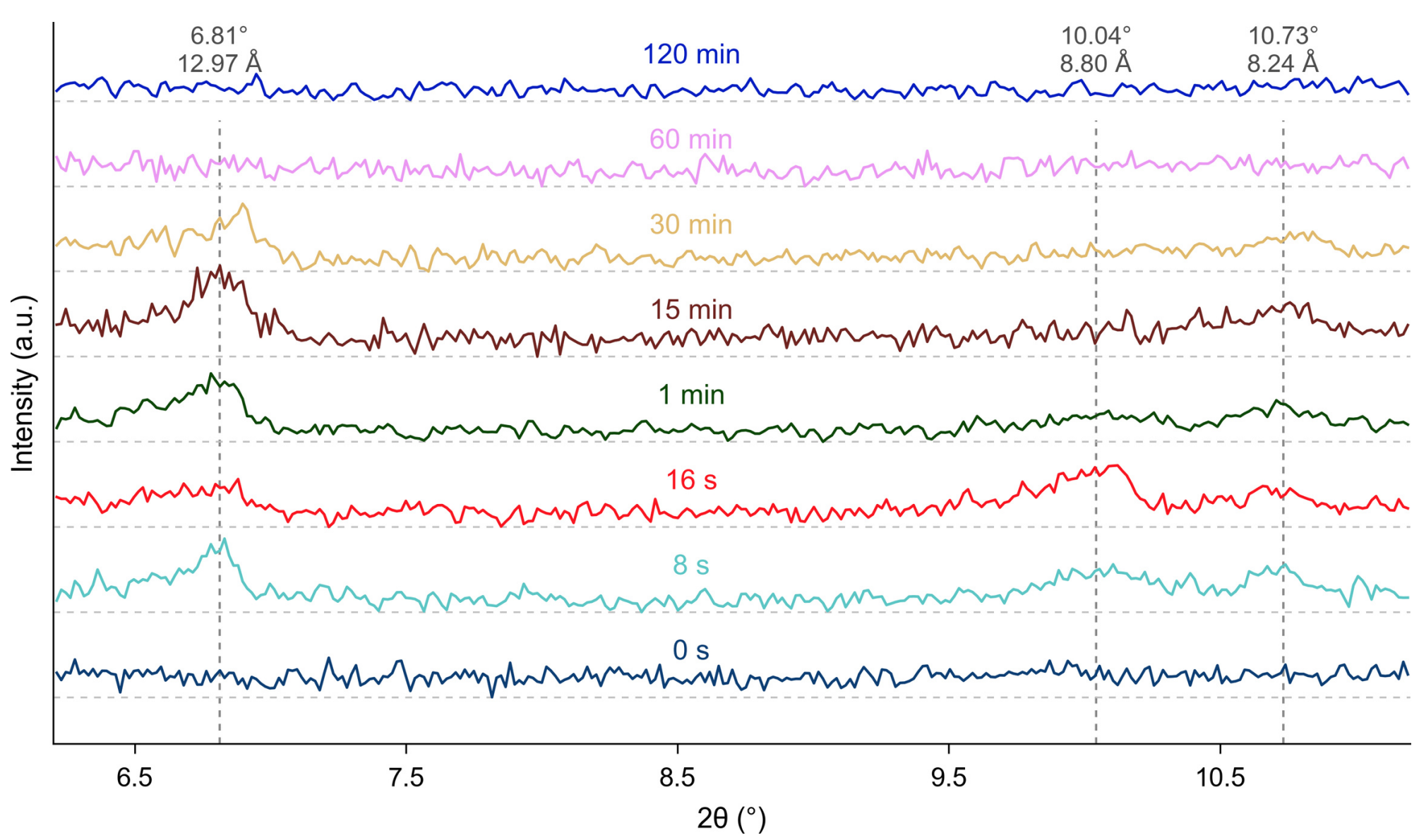

**Supplementary Fig. 21 | Transient low-angle reflections identify crystalline intermediate phases.**

Vertically offset ex situ XRD patterns of the WBG stacks after the indicated annealing times. Reflections at 6.81°, 10.04° and 10.73° (d = 12.97, 8.80 and 8.24 Å) appear within 8 s and disappear by 60 min, coincident with growth of the 3D perovskite reflection.

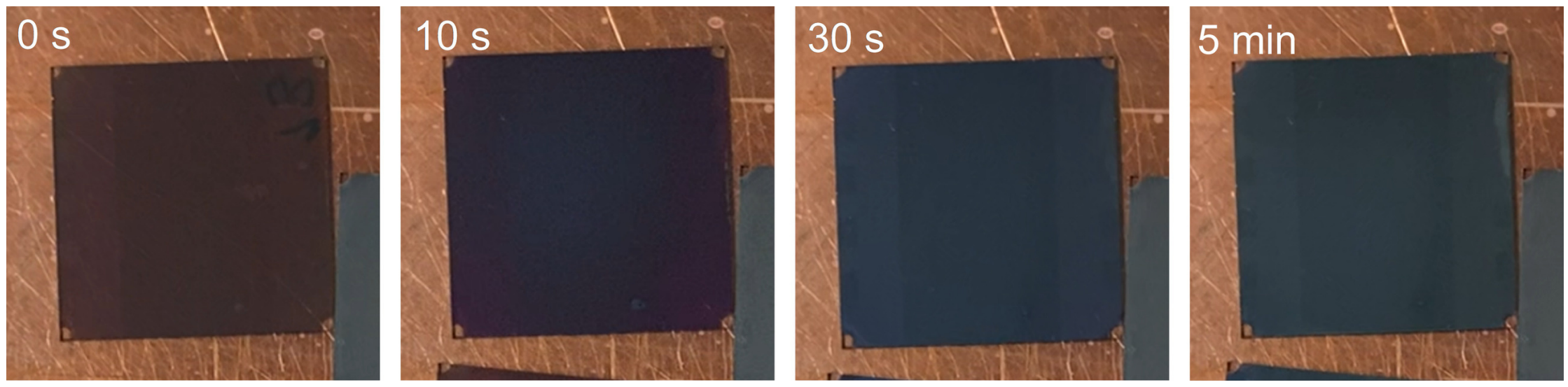

**Supplementary Fig. 22 | Rapid darkening of Sn–Pb NBG stacks.**

Time-series photographs of $SnI_2/PbI_2/FAI$ stacks annealed in an $N_2$-filled glovebox at 150 °C. The stack develops a dark appearance within the first seconds. The corresponding recording is Supplementary Video 3.

**Supplementary Video 3 | Rapid darkening of Sn–Pb NBG sTE stacks during annealing.** Time-lapse video of a sequentially evaporated $SnI_2/PbI_2/FAI$ NBG precursor stack during annealing from 0 to 5 min. The stack rapidly develops a dark perovskite-like appearance during early-stage annealing.

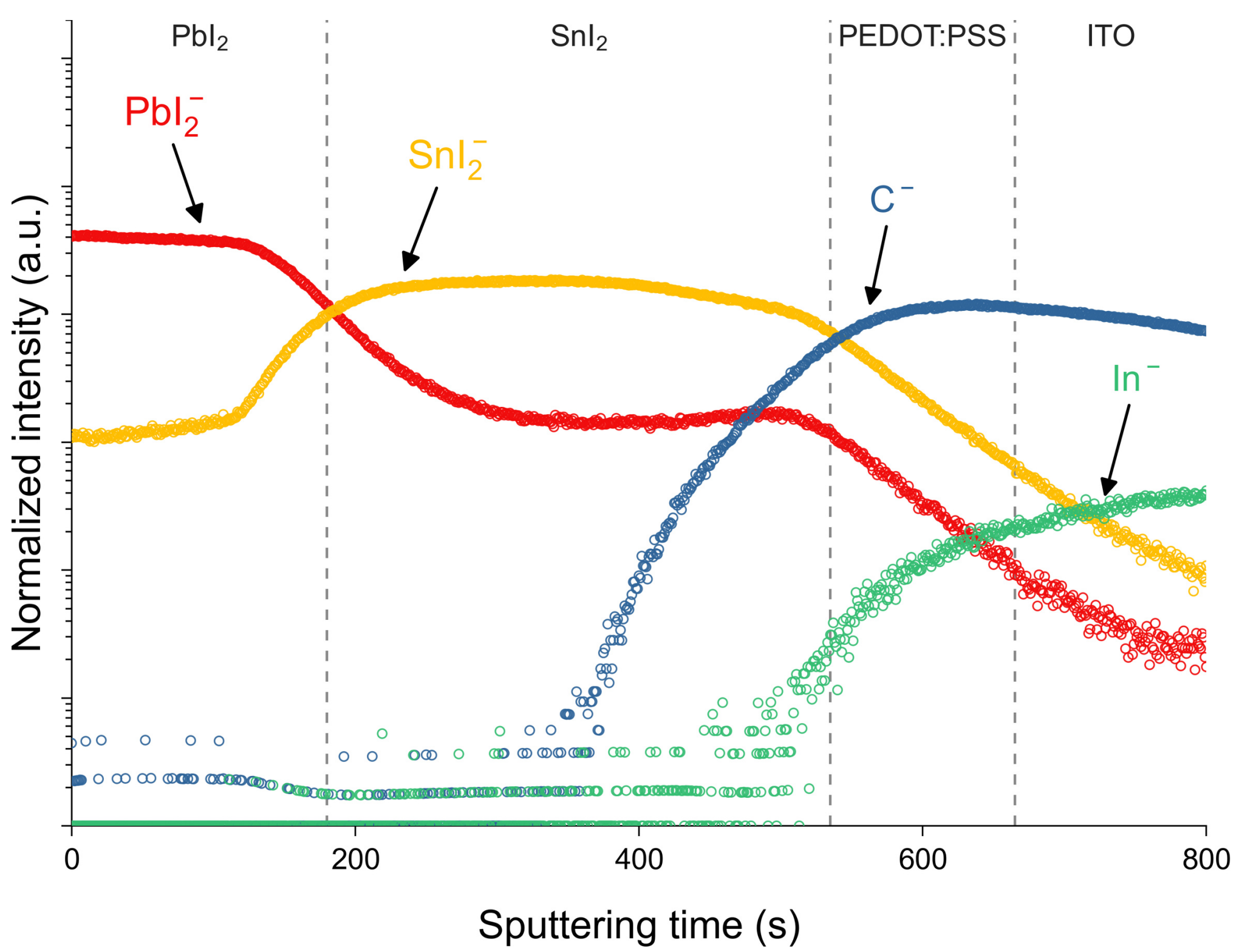

**Supplementary Fig. 23 | The NBG inorganic template is chemically stratified.**

ToF-SIMS depth profiles of the as-deposited ITO/PEDOT:PSS/$SnI_2$/$PbI_2$ template. $PbI_2^-$ and $SnI_2^-$ track the Pb- and Sn-rich precursor regions, while $C^-$ and $In^-$ mark PEDOT:PSS and ITO. Dashed lines indicate the apparent transitions between chemically distinct layers.

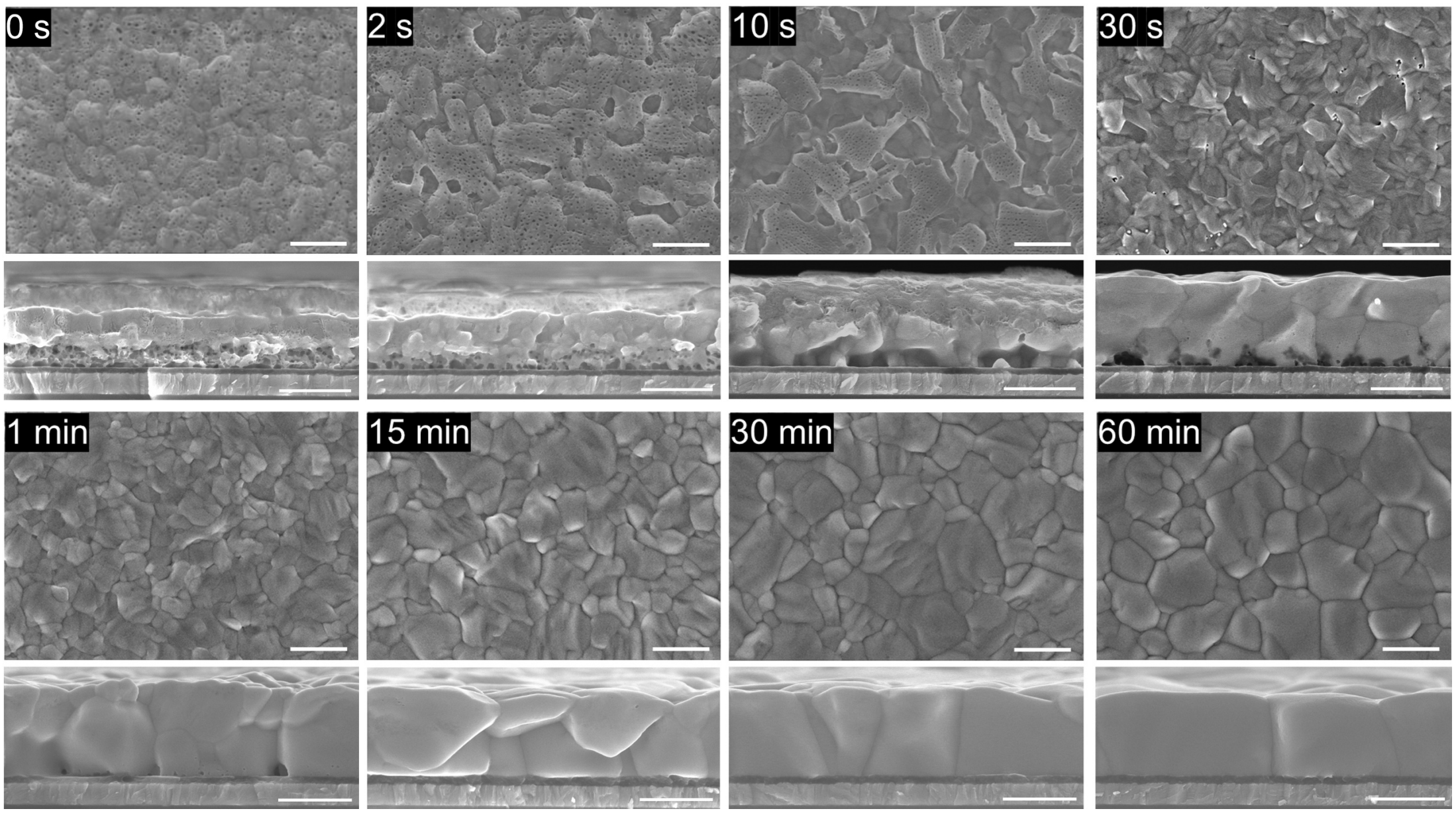

**Supplementary Fig. 24 | Morphological evolution of NBG stacks during annealing.**

Top-view and cross-sectional SEM images of ITO/PEDOT:PSS/$SnI_2$/$PbI_2$/FAI stacks after the indicated annealing times; each image was acquired from a separate sample. Scale bars, 1 μm (top views) and 500 nm (cross-sections).

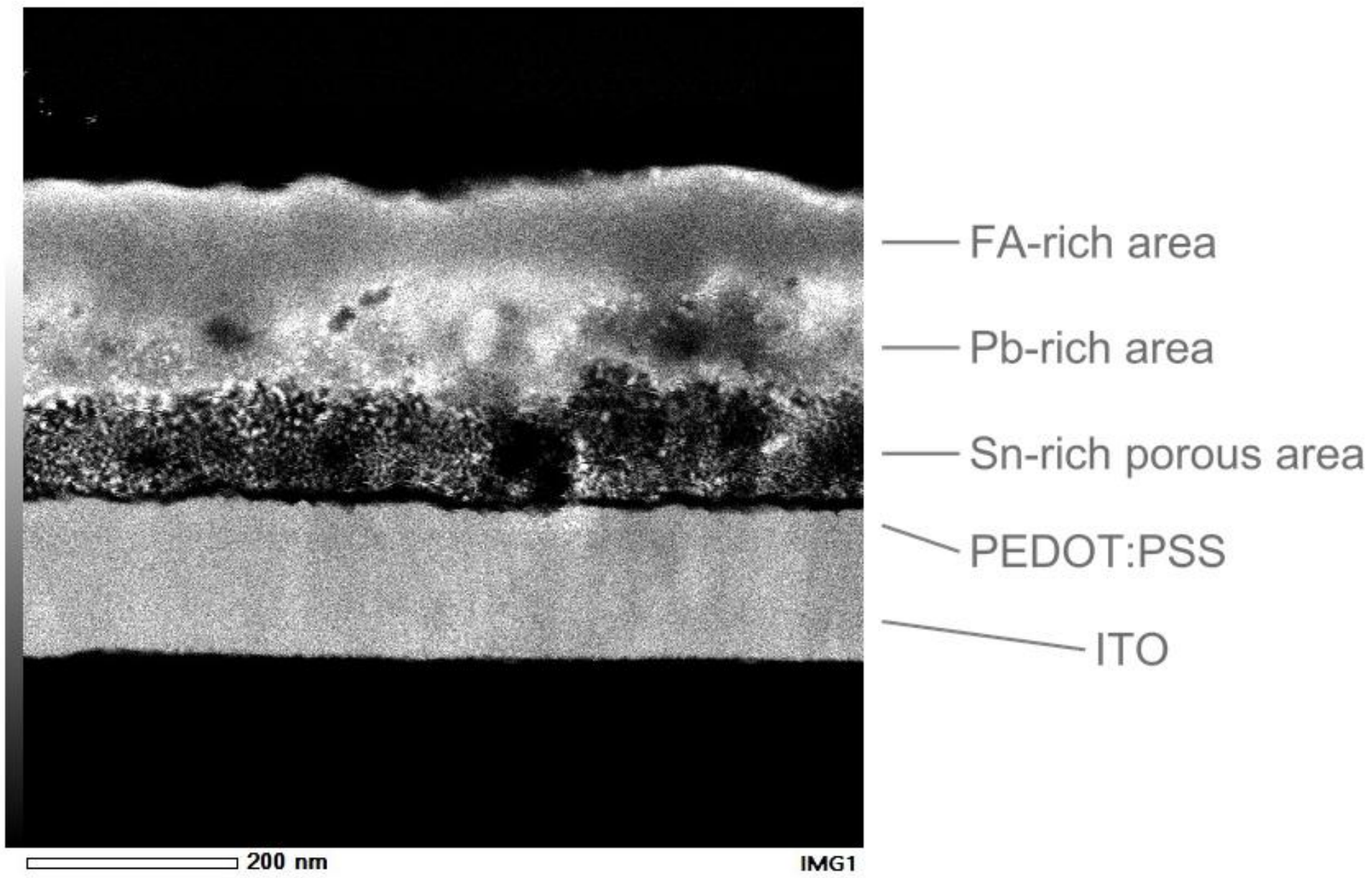

**Supplementary Fig. 25 | Layered architecture and buried porosity of the as-deposited NBG stack.**

Cross-sectional HAADF-STEM image of ITO/PEDOT:PSS/$SnI_2$/$PbI_2$/FAI after FAI deposition and before annealing. The image resolves three morphologically distinct regions; by comparison with the ToF-SIMS assignments, these correspond to an FA-rich upper region, a Pb-rich middle region and a porous Sn-rich region above PEDOT:PSS. Scale bar, 200 nm.

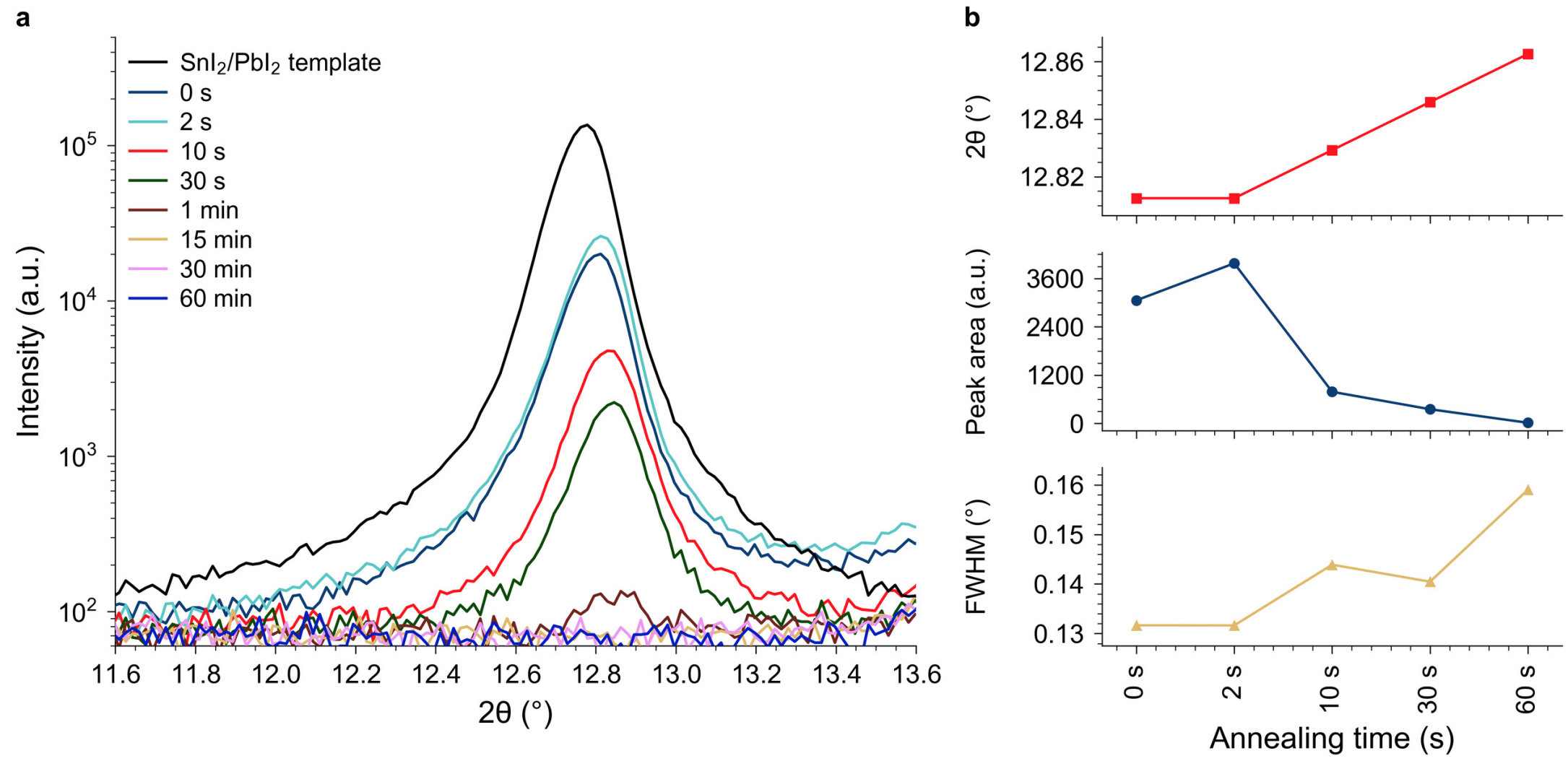

**Supplementary Fig. 26 | Rapid consumption of the inorganic precursor in the NBG stack.**

**a,** Ex situ XRD patterns of the $SnI_2/PbI_2$ inorganic template and full NBG stacks in the 11.6–13.6° 2θ range after the indicated annealing times. The precursor-related reflection near 12.8° decreases rapidly and is nearly undetectable after 1 min. **b,** Measured peak position and fitted integrated area and FWHM of the same inorganic precursor-related diffraction feature during the first 60 s.

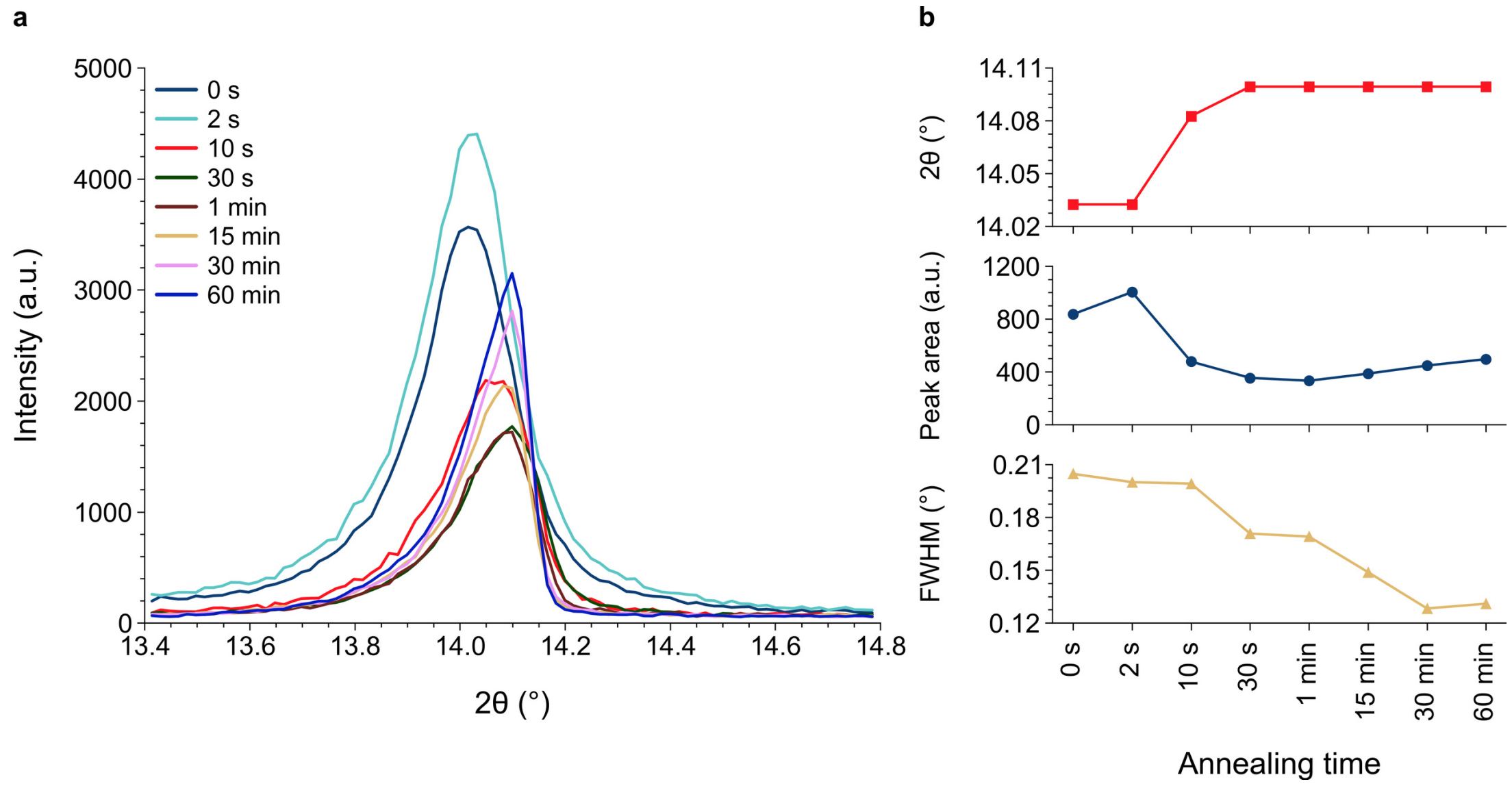

**Supplementary Fig. 27 | Evolution of the NBG perovskite diffraction feature.**

**a,** Ex situ XRD patterns of ITO/PEDOT:PSS/$SnI_2$/$PbI_2$/FAI stacks in the 13.4–14.8° 2θ range after the indicated annealing times. **b,** Measured peak position and fitted integrated area and FWHM of the same perovskite-related diffraction feature. The reflection is already present before annealing, shifts to higher 2θ and loses integrated area during short annealing, and then recovers in area and narrows at a nearly stabilized peak position during prolonged annealing.

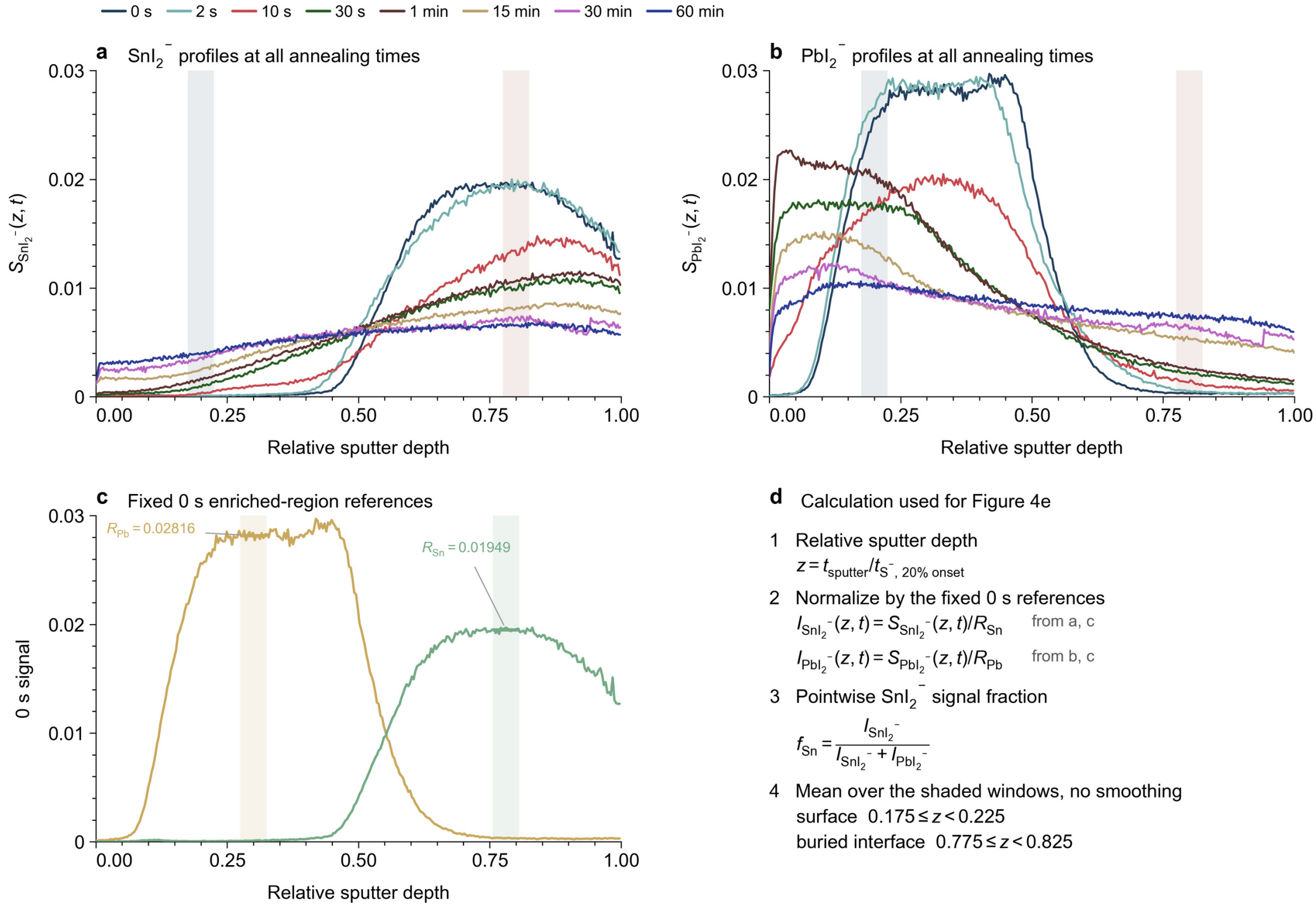

**Supplementary Fig. 28 | Depth-resolved $SnI_2^-$ and $PbI_2^-$ signals and the $SnI_2^-$ signal-fraction analysis.**

Negative-ion ToF-SIMS depth profiles of ITO/PEDOT:PSS/$SnI_2$/$PbI_2$/FAI stacks after the indicated annealing times. **a, b,** $SnI_2^-$ and $PbI_2^-$ profiles plotted against relative sputter depth. For each annealing time, the relative sputter depth, $z$, was obtained by dividing the sputter time by the 20%-of-maximum rising-edge onset time of $S^-$. Shaded regions indicate the sampling windows centred at $z = 0.20$ and $z = 0.80$. **c,** The 0 s $PbI_2^-$ and $SnI_2^-$ profiles used to define the reference signals $R_{Pb}$ and $R_{Sn}$, respectively. $R_{Pb}$ was obtained from the $PbI_2$-rich region at $0.275 \le z < 0.325$, whereas $R_{Sn}$ was obtained from the $SnI_2$-rich region at $0.756 \le z < 0.806$. **d,** The normalized signals were calculated as $I_{SnI_2^-} = \frac{S_{SnI_2^-}}{R_{Sn}}, I_{PbI_2^-} = \frac{S_{PbI_2^-}}{R_{Pb}}$, and the $SnI_2^-$ signal fraction as $f_{Sn} = \frac{I_{SnI_2^-}}{I_{SnI_2^-}+I_{PbI_2^-}}$. The values shown in **Fig. 4e** are the means of the pointwise fractions within $0.175 \le z < 0.225$ on the surface side and $0.775 \le z < 0.825$ on the buried-interface side. No smoothing was applied.

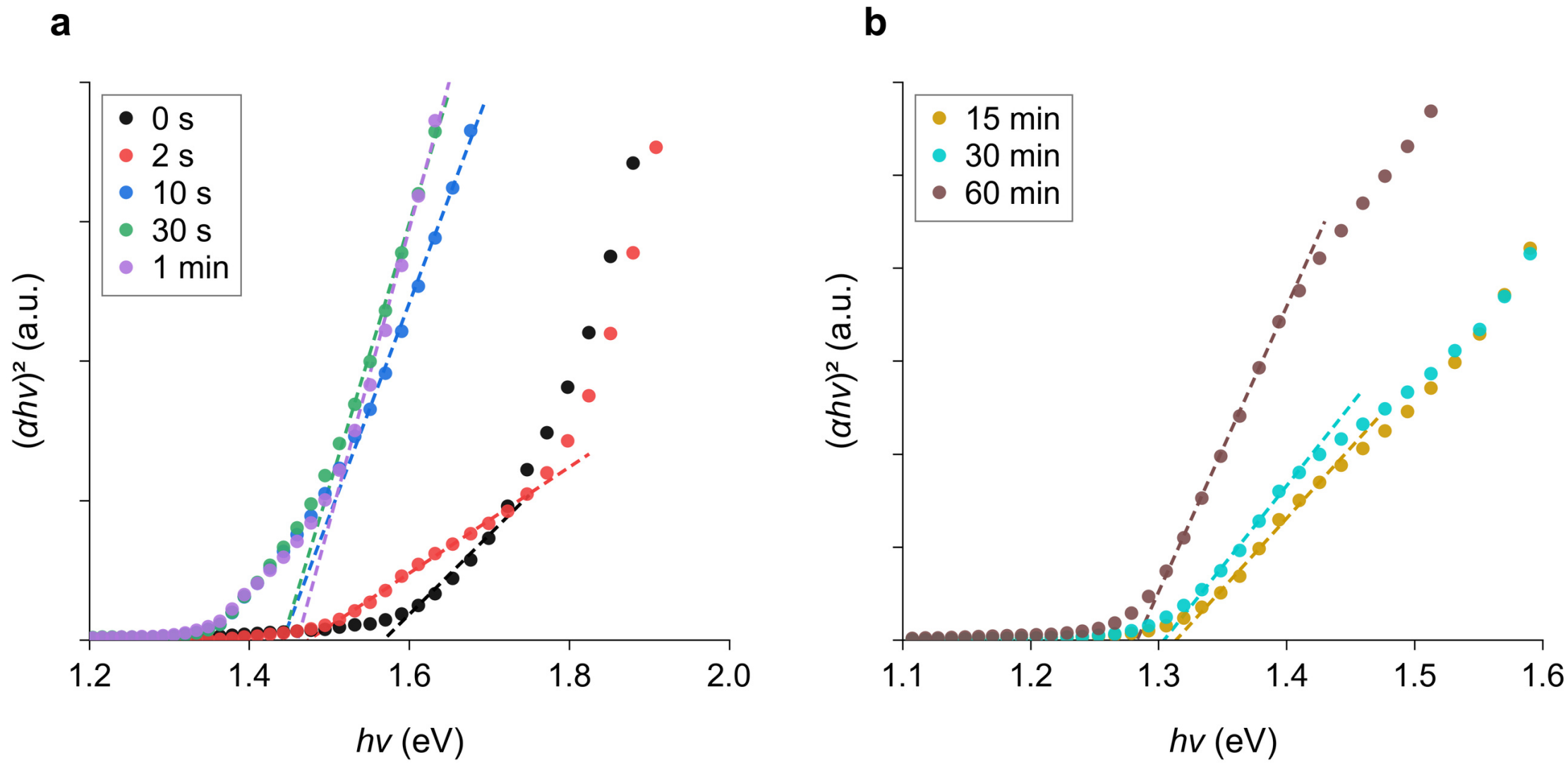

**Supplementary Fig. 29 | Progressive narrowing of the NBG optical bandgap.**

Direct-transition Tauc plots of the ITO/PEDOT:PSS/$SnI_2$/$PbI_2$/FAI stacks after the indicated annealing times; dashed lines are linear fits to the absorption edges. a, Early conversion from 0 s to 1 min. A high-energy absorption edge is present before annealing and shifts rapidly to below 1.5 eV after seconds of annealing. **b,** Prolonged annealing from 15 to 60 min further narrows the apparent bandgap to below 1.3 eV.

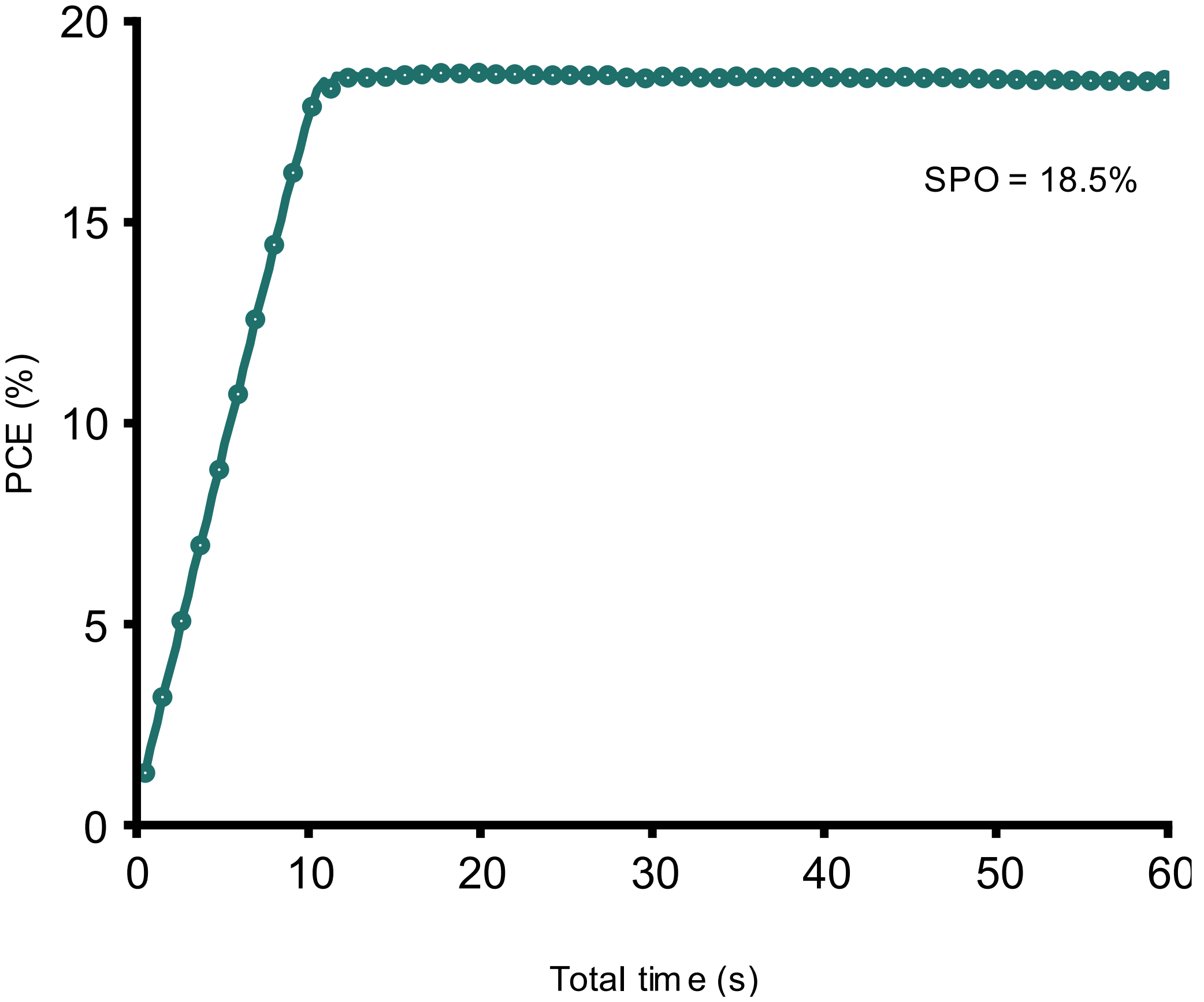

**Supplementary Fig. 30 | Stabilized power output (SPO) of the evaporated tandem solar cell.**

Power conversion efficiency during 60 s of maximum-power-point tracking under simulated AM 1.5G illumination. The output stabilizes at 18.5%. The device was measured through a 0.0616-$cm^2$ aperture.

**Supplementary Table 1 | Precursor layer thicknesses and photovoltaic performance of perovskite solar cells with sTE absorbers.** Nominal thicknesses of the sequentially thermally evaporated $SnI_2$, $PbBr_2$, $PbI_2$, formamidinium iodide (FAI) and CsI are listed together with the corresponding device parameters ($J_{SC}$, $V_{OC}$, FF and PCE) extracted from current–voltage measurements under simulated AM 1.5G illumination (100 mW cm$^{-2}$). The bandgap ($E_g$) was determined from the inflection point of the EQE spectra. "/" indicates that the corresponding layer was not deposited.

| $SnI_2$ (nm) | $PbBr_2$ (nm) | $PbI_2$ (nm) | FAI (nm) | CsI (nm) | $J_{SC}$ (mA cm$^{-2}$) | $V_{OC}$ (V) | FF (%) | PCE (%) | $E_g$ from EQE (eV) |
|---|---|---|---|---|---|---|---|---|---|
| 110 | / | 110 | 480 | / | 26.83 | 0.784 | 74.92 | 15.77 | 1.26 |
| / | 10 | 220 | 350 | 11 | 22.91 | 1.065 | 82.04 | 20.01 | 1.54 |
| / | 50 | 220 | 350 | 11 | 21.00 | 1.110 | 82.19 | 19.16 | 1.61 |
| / | 100 | 220 | 350 | 11 | 19.49 | 1.171 | 84.74 | 19.34 | 1.70 |
| / | 150 | 60 | 290 | 11 | 17.46 | 1.215 | 79.93 | 16.95 | 1.78 |
| / | 250 | 50 | 350 | 11 | 14.08 | 1.206 | 78.36 | 13.31 | 1.90 |
| / | 250 | 0 | 350 | 11 | 12.98 | 1.276 | 81.42 | 13.49 | 1.96 |

**Supplementary Table 2 | Nominal composition of the sTE precursor stacks in Fig. 2f.** Layer thicknesses of the sequentially thermally evaporated $PbBr_2$, $PbI_2$, formamidinium iodide (FAI) and CsI are nominal values. The nominal Br/(Br + I) ratio was calculated by converting each thickness into a molar amount using the densities of $PbBr_2$ (6.66 g cm$^{-3}$), $PbI_2$ (6.16 g cm$^{-3}$), CsI (4.51 g cm$^{-3}$) and FAI (2.48 g cm$^{-3}$),[34,35] counting two halide ions per $PbX_2$ and one per FAI or CsI. T and B label ternary and binary stacks, ordered by nominal Br/(Br + I). "/" indicates that the corresponding layer was not deposited.

| Label | $PbBr_2$ (nm) | $PbI_2$ (nm) | FAI (nm) | CsI (nm) | Nominal Br/(Br + I) (%) | $E_g$ (eV) |
|---|---|---|---|---|---|---|
| *Ternary sTE stacks ($PbBr_2$/$PbI_2$/FAI/CsI)* | | | | | | |
| T1 | 100 | 140 | 480 | 11 | 25.1 | 1.70 |
| T2 | 140 | 140 | 480 | 11 | 31.9 | 1.74 |
| T3 | 140 | 100 | 480 | 11 | 34.2 | 1.77 |
| T4 | 140 | 60 | 480 | 11 | 36.8 | 1.77 |
| T5 | 180 | 140 | 480 | 11 | 37.6 | 1.73 |
| T6 | 180 | 60 | 560 | 11 | 39.9 | 1.76 |
| T7 | 140 | 20 | 480 | 11 | 39.9 | 1.80 |
| T8 | 140 | 10 | 480 | 11 | 40.8 | 1.79 |
| T9 | 140 | 5 | 480 | 11 | 41.2 | 1.81 |
| T10 | 180 | 60 | 520 | 11 | 41.3 | 1.78 |
| T11 | 220 | 140 | 480 | 11 | 42.4 | 1.78 |
| T12 | 180 | 60 | 480 | 11 | 42.9 | 1.81 |
| *Binary sTE stacks ($PbBr_2$/FAI/CsI)* | | | | | | |
| B1 | 140 | / | 600 | 11 | 36.5 | 1.83 |
| B2 | 140 | / | 560 | 11 | 38.1 | 1.85 |
| B3 | 140 | / | 520 | 11 | 39.8 | 1.85 |
| B4 | 140 | / | 480 | 11 | 41.7 | 1.85 |
| B5 | 180 | / | 600 | 11 | 42.5 | 1.87 |
| B6 | 180 | / | 560 | 11 | 44.2 | 1.89 |
| B7 | 160 | / | 480 | 11 | 45.0 | 1.87 |
| B8 | 180 | / | 520 | 11 | 46.0 | 1.91 |
| B9 | 180 | / | 480 | 11 | 47.9 | 1.89 |
| B10 | 200 | / | 480 | 11 | 50.5 | 1.91 |